\documentclass[12pt]{article}

\usepackage{graphicx}
\usepackage{epsfig}

\newcommand{\Z}{{\sf Z \!\!\! Z}}

\newcommand{\Sign}{\mbox{Sign}}
\newcommand{\cxu}{c_{x \uparrow}}
\newcommand{\cxd}{c_{x \downarrow}}
\newcommand{\nxu}{n_{x \uparrow}}
\newcommand{\nxd}{n_{x \downarrow}}
\newcommand{\cyu}{c_{y \uparrow}}
\newcommand{\cyd}{c_{y \downarrow}}
\newcommand{\nyu}{n_{y \uparrow}}
\newcommand{\nyd}{n_{y \downarrow}}
\makeatletter
\@addtoreset{equation}{section}
\makeatother

\begin{document}
 
\title{Meron-Cluster Approach to Systems of \\ Strongly Correlated Electrons}
\author{S. Chandrasekharan$^a$, J. Cox$^b$, J.C. Osborn$^a$\thanks
{Current address Universiry of Utah, Salt Lake City, USA} ~and 
U.-J. Wiese$^{b,c}$
\\ \\
$^a$ Department of Physics, Box 90305, Duke University, \\
Durham, North Carolina 27708 \\
$^b$ Center for Theoretical Physics, Laboratory for Nuclear Science \\
and Department of Physics, Massachusetts Institute of Technology, \\
Cambridge, Massachusetts 02139 \\
$^c$ Institute for Theoretical Physics, Bern University, \\
CH-3012, Switzerland}

\maketitle

\begin{abstract} 
Numerical simulations of strongly correlated electron systems suffer
from the notorious fermion sign problem which has prevented progress
in understanding if systems like the Hubbard model display
high-temperature superconductivity. Here we show how the fermion sign
problem can be solved completely with meron-cluster methods in a large
class of models of strongly correlated electron systems, some of which
are in the extended Hubbard model family and show s-wave
superconductivity. In these models we also find that on-site repulsion
can even coexist with a weak chemical potential without introducing
sign problems. We argue that since these models can be simulated
efficiently using cluster algorithms they are ideal for studying many
of the interesting phenomena in strongly correlated electron systems.
\end{abstract}

\newpage

\section{Introduction}

Strongly correlated electrons are among the most interesting condensed matter
systems. In particular, the doping of anti-ferromagnetic cuprate layers leads 
to high-temperature superconductivity. Understanding the dynamics of strongly 
correlated electron systems is very challenging because perturbative analytic 
techniques fail for systems of many strongly coupled degrees of freedom. 
On the other hand, numerical simulation methods are applicable in these cases.
Unfortunately, numerical simulations of models involving fermions are far from 
trivial. A standard technique is to linearize four-fermion interactions by the 
introduction of bosonic fields \cite{Whi89}. Then the fermions are integrated 
out, leaving behind a fermion determinant which acts as a non-local effective 
action for the bosonic fields. In some cases, for example, in undoped cuprate 
layers at half-filling, the fermion determinant is positive and can be 
interpreted as a probability for the bosonic field configurations. Then 
standard importance sampling techniques can be applied to the bosonic theory. 
Still, due to the non-locality of the effective action, such simulations are 
very time consuming. In other cases of physical interest --- in particular for 
doped cuprates away from half-filling --- the fermion determinant may become 
negative and can hence not be interpreted as a probability for the bosonic 
configurations. When the sign of the fermion determinant is included in 
measured observables, the fluctuations in the sign give rise to dramatic 
cancellations. As a consequence, the relative statistical errors of observables
grow exponentially with the volume and the inverse temperature of the system. 
This makes it impossible in practice to simulate large systems in the 
low-temperature limit. In particular, the fermion sign problem prevents 
numerical simulations of the Hubbard model away from half-filling and is the 
major stumbling block against understanding high-temperature superconductivity 
with numerical methods.

Dealing with the fermion sign problem after the fermions have been integrated
out is incredibly complicated since the resulting bosonic
effective action is non-local. Before integrating them out, fermions are
usually described by anti-commuting Grassmann variables which are practically
impossible to simulate directly. Here we propose a different strategy. Instead 
of working with Grassmann coherent states, we formulate the fermionic path 
integral in the Fock state basis of the Hilbert space. Then a fermion 
configuration is described in terms of bosonic occupation numbers which define 
fermion world-lines. In this case, the bosonic occupation numbers interact 
locally. However, besides the positive bosonic Boltzmann weight, there is a 
fermion permutation sign that results from the Pauli principle. Two fermions 
that interchange their positions during the Euclidean time-evolution give rise 
to a minus-sign. In general, the fermion world-lines which are periodic in 
Euclidean time define a permutation of fermion positions. The sign of this 
permutation is the fermion sign. The resulting sign problem is often more 
severe than the one associated with the fermion determinant. For example, in 
the world-line formulation a sign problem arises for the Hubbard model even at 
half-filling, while the fermion determinant is positive in that case. 

Fortunately, in the Fock state basis the sign problem can be
completely eliminated, at least in a large (but still restricted)
class of models.  This was first demonstrated for a system of spinless
lattice fermions in \cite{Cha99,Cha00} by re-writing the partition
function in terms of the statistical mechanics of closed loops with
non-negative Boltzmann weights.  Loops with a certain topology, referred
to as {\em merons}, do not contribute to the partition function. A
local algorithm was constructed in the loop space which avoided
meron-clusters and thus solved the sign problem.  It is well known
that spinless fermions can be converted to relativistic staggered
fermions by introducing carefully chosen phase factors with every
fermion hop \cite{Suss77}. Using this strategy, the solution to the
fermion sign problem with spinless fermions has been exploited to
extensively study the critical behavior of a second order chiral phase
transition in the universality class of the Ising model
\cite{Cha00a,Cha00b,Cox00}.

Here we extend these successful techniques to fermions with spin and, in 
particular, to systems in the Hubbard model family. Using a specific
example, we show how one can build models so that the sign problem is
completely solved. For technical reasons related to the details of the 
cancellation of signs, the method does not apply to the standard Hubbard 
model Hamiltonian. Additional terms must be 
included. Still, the modifications do not affect the symmetries of the 
problem and we believe that the physics of some of the models discussed 
here is qualitatively similar to the standard Hubbard model. It should be 
pointed out that there is no reason to concentrate on the standard Hubbard 
model except for the simple form of its Hamiltonian. The fact that 
for similar Hamiltonians our method can completely eliminate the sign 
problem and yield a very efficient fermion algorithm --- which is not the 
case for the standard Hubbard model --- is reason enough to replace the 
standard Hubbard model by a modified Hamiltonian. After all, we want to 
focus on understanding the general physical phenomena underlying 
high-temperature superconductivity --- not necessarily on the details
of one particular 
model Hamiltonian. Interestingly, in the modified model the sign problem 
can be eliminated even in the presence of an on-site repulsion at a finite 
chemical potential as long as a certain inequality is satisfied. 

\subsection{The Fermion Sign Problem}

Let us first discuss the nature of the fermion sign problem. We consider a 
fermionic path integral
\begin{equation}
Z_f = \sum_{[n]} \Sign[n] \exp(-S[n])
\end{equation}
over configurations of occupation numbers $n$ with a Boltzmann weight 
whose magnitude is $\exp(-S[n])$ and sign is $\Sign[n] = \pm 1$. Here 
$S[n]$ is the action of a corresponding bosonic model with partition 
function
\begin{equation}
Z_b = \sum_{[n]} \exp(-S[n]).
\end{equation}
A fermionic observable $O[n]$ is obtained in a simulation of the bosonic 
ensemble as
\begin{equation}
\langle O \rangle_f = \frac{1}{Z_f} \sum_{[n]} O[n] \Sign[n] \exp(-S[n]) =
\frac{\langle O \ \Sign \rangle}{\langle \Sign \rangle}.
\end{equation}
The average sign in the simulated bosonic ensemble is given by
\begin{equation}
\langle \Sign \rangle = \frac{1}{Z_b}{\sum_{[n]} \Sign[n] \exp(-S[n])} = 
\frac{Z_f}{Z_b} = \exp(- \beta V \Delta f).
\end{equation}
The expectation value of the sign is exponentially small in both the volume $V$
and the inverse temperature $\beta$ because the difference between the free 
energy densities $\Delta f = f_f - f_b$ of the fermionic and bosonic systems is
always positive. Hence, the fermionic expectation value $\langle O \rangle_f$
--- although of order one --- is measured as the ratio of two exponentially 
small signals which are very hard to extract from the statistical noise. This 
is the origin of the sign problem. We can estimate the statistical error of the
average sign in an ideal simulation of the bosonic ensemble which generates $N$
completely uncorrelated configurations as
\begin{equation}
\frac{\sigma_{\mathrm{Sign}}}{\langle \Sign \rangle} = 
\frac{\sqrt{\langle \Sign^2 \rangle - \langle \Sign \rangle^2}}
{\sqrt{N} {\langle \Sign \rangle}} = \frac{\exp(\beta V \Delta f)}{\sqrt{N}}.
\end{equation}
The last equality results from taking the large $\beta V$ limit and using
$\Sign^2 = 1$. In order to determine the average sign with sufficient
accuracy one needs
on the order of $N = \exp(2 \beta V \Delta f)$ measurements. For large volumes 
and small temperatures this is impossible in practice. 

In the meron-cluster approach one solves the sign problem in two 
steps. In the first step one analytically re-writes the partition function
using new variables such that it is possible to exactly cancel 
all negative weight configurations with configurations of positive weights.
This group of configurations then does not contribute to the partition 
function. 
The remaining configurations are guaranteed to be positive. Thus, in the new 
variables, one effectively obtains Boltzmann weights with $\Sign = 0,1$. 
At this stage, despite the fact that all negative signs have been
eliminated, only one 
half of the sign problem has been solved since an algorithm that naively
generates configurations with $\Sign = 0$ or $1$ generates an exponentially
large number of zero weight configurations. Thus, it is important to introduce
a second step in which one avoids configurations that have been canceled
analytically. In practice it is often useful to allow these 
zero-weight configurations to occur in a controlled manner since these 
configurations may contribute to observables. In a numerical algorithm
a local Metropolis 
decision ensures that contributions of $0$ and $1$ occur with similar 
probabilities. This solves the sign problem completely.

\subsection{The Meron-Cluster Method}

\label{mcm}

The central idea behind the meron-cluster method is to express the 
partition function, which is originally written as a sum over
weights of fermion occupation number configurations, as a sum over 
weights of configurations which contain both fermions and new bond 
variables. Mathematically this means
\begin{equation}
\label{pfnb}
Z_f = \sum_{[n]} \Sign[n] \exp(-S[n]) = \sum_{[n,b]} \Sign[n,b] 
\exp(-S[n,b])
\end{equation}
where the bond variables $b$ carry information about whether two lattice 
sites are connected or not. A cluster is defined as a set of
connected lattice sites whose flip exchanges occupied and empty sites, 
i.e., the occupied sites on that cluster are emptied and the originally 
empty sites now become occupied. This step of re-writing the partition
function is well known and has been used in earlier attempts of 
constructing fermion cluster algorithms as discussed in \cite{Wie93} and 
\cite{Amm98}. The recent progress results from the observation that the sign 
problem contained in a partition function of the form (\ref{pfnb}) is 
completely solvable if the magnitude $W[n,b]$ and the sign $\Sign[n,b]$
of the Boltzmann weight satisfy three requirements:
\begin{enumerate}
\item The magnitude of the Boltzmann weight $\exp(-S[n,b])$ does not
change when any cluster is flipped.
\item The effect of a cluster-flip on the sign of the Boltzmann 
weight $\Sign[n,b]$ is independent of the orientation of all other
clusters. 
\item Starting from any configuration $[n,b]$, it must be possible to
flip clusters and reach a {\em reference} configuration $[n_{\rm{ref}},b]$, 
which is guaranteed to have a positive Boltzmann weight. 
\end{enumerate}
In addition, a formula for the change in the sign of a configuration
due to a cluster-flip has been derived recently \cite{Cha00c}.
Using this formula and other tricks it is usually possible to satisfy 
the first two properties for any given model. Restrictions in the class 
of solvable models arise due to the inability to satisfy the third property 
listed above,
i.e., the existence of a reference configuration. However, we have
been able to construct useful reference configurations for a variety
of models.

The essential consequence of the three basic properties necessary for
the meron-cluster approach to work is that clusters can be
characterized by their effect on the sign. Clusters whose flip changes
the sign are referred to as merons. Clearly a configuration with a
meron-cluster contributes zero to the partition function after a
partial re-summation over cluster-flips is performed\footnote{This
partial re-summation over cluster-flips is often referred to as an
improved estimator.}. All configurations without any merons contribute
positively to the partition function.  Thus the partition function can
be re-written in terms of positive semi-definite Boltzmann weights,
i.e., $\Sign = 0,1$. In order to avoid the zero-weight configurations,
a re-weighting method that eliminates configurations containing
multiple meron-clusters is included. In order to perform
this re-weighting, all observables must also be measured using improved
estimators. Fortunately, this is possible for most quantities of
physical interest. Usually, physically interesting observables
receive non-zero contributions also from configurations containing merons. For
example, vacuum condensates receive contributions from the one-meron
sector whereas observables derived from certain two-point correlation
functions receive contributions from zero- and two-meron configurations
only. Four-point correlation functions also receive non-zero
contributions from four-meron configurations. Fortunately, one is
usually not interested in multi-point correlation functions and for most
purposes one can completely eliminate the configurations with more
than two merons. This is exactly what happens in the re-weighting step
of the meron-cluster algorithm.  In this paper we do not discuss the
re-weighting step although this is an essential part of the complete
solution of the sign problem. We assume that once all negative signs
are eliminated,
one can design a re-weighting step that does not generate exponentially 
long auto-correlation times.

Although the three requirements for solving the sign problem
with the meron-cluster method are somewhat restrictive, a large class
of Hamiltonians for strongly interacting fermions is consistent with
them. For example, in the case of spinless fermions it is necessary to
include a certain amount of nearest-neighbor fermion repulsion
\cite{Cha99} or a large chemical potential \cite{Cha00} depending on
the hopping term.  As we will see, similar restrictions arise for
fermions with spin. In particular, the standard Hubbard model
Hamiltonian cannot be treated with our method, while slightly modified
models in the Hubbard model family can. In this paper we construct a
large class of $SU(2)$ spin symmetric Hamiltonians for spin one-half
fermions with nearest-neighbor interactions to which the meron-cluster
method can be applied. The numerical results for one such model 
related to s-wave superconductivity have been discussed elsewhere \cite{Cha01}.

The meron-cluster idea was first developed for a bosonic model with 
a complex action --- the 2-d $O(3)$ model at non-zero vacuum-angle $\theta$ 
\cite{Bie95}. The term meron was introduced first to denote half-instantons. 
Indeed the meron-clusters in the 2-d $O(3)$ model are half-instantons 
\cite{Bie95}. The instanton number is the integer-valued topological 
charge $Q \in \Z$, which gives rise to a complex
phase $\exp(i \theta Q)$ that contributes to the Boltzmann factor. At 
$\theta = \pi$ the phase factor reduces to a sign $\exp(i \pi Q) = (-1)^Q$, 
which distinguishes between sectors with even and odd topological charges. In 
the 2-d $O(3)$ model the meron-clusters carry half-integer topological charges.
When such a cluster is flipped, its topological charge changes sign and hence
the total topological charge of the configuration changes by an odd integer. 
Thus, the phase factor $(-1)^Q$ changes sign when a meron-cluster is flipped.
Similarly, a meron-flip in the fermion cluster algorithm changes the
sign of a fermion world-line configuration. Interestingly a $\Z(2)$ 
topological number can be assigned to each fermionic configuration. Any 
configuration with $\Sign = 1$ is topologically trivial because it 
represents an even permutation of fermions. A configuration with 
$\Sign = - 1$ is topologically distinct, because an odd permutation cannot 
be changed into an even one by a continuous deformation of fermion 
world-lines. In this sense, configurations with $\Sign = - 1$ are instantons. 
Flipping any meron changes the fermion sign and hence unwinds the instanton. 
In two spatial dimensions one can extend the analogy to the 2-d $O(3)$ model 
even further. Then particles can exist with any statistics, characterized by 
a parameter $\theta$ that varies between $\theta = 0$ for bosons and 
$\theta = \pi$ for fermions. For anyons the fermion sign is replaced by a 
complex phase factor $\exp(i \theta H)$, where $H \in Z$ is the integer-valued 
Hopf number. In that case the merons are half-Hopf-instantons. 

The merons are not just of algorithmic interest. In fact, they are physical 
objects that allow us to 
understand the topology of the fermion world-lines and hence the origin of the
fermion sign. The merons effectively bosonize a fermionic theory. When a 
fermionic model is formulated in terms of bosonic occupation numbers which 
interact locally, the fermion sign still induces non-local topological 
interactions. Hence, the resulting theory is not really bosonized. The merons,
however, decompose the global topology of the fermion world-lines into
manageable contributions that are local on the scale of the correlation length.
This allows us to simulate fermions almost as efficiently as bosons.

\subsection{Organization of the Paper}

The paper is organized as follows. In section 2 the essential ideas
behind the meron-cluster approach are illustrated for the simple case of 
spinless fermions. We discuss how the three properties necessary
for the complete solution to the sign problem, discussed in
section \ref{mcm}, place restrictions on the class of models that are 
solvable with the meron-cluster algorithm. 
In section 3 we build a class of models of fermions with spin
which are solvable using meron-cluster methods. We also discuss improved 
estimators for some physical quantities useful for the study of 
superconductivity that receive contributions from zero- and two-meron 
configurations and present some data from a recent numerical simulation.
We also show how one can extend these models by adding new terms to the
Hamiltonian that appear to violate the properties necessary for the 
meron-cluster method to work but do not introduce sign problems. 
In particular, we show how a repulsive Hubbard model becomes solvable 
in the presence of a chemical potential. Section 4 contains a discussion
of the efficiency of meron-cluster algorithms. Finally, in section 5 we end 
with some conclusions.

\section{Meron-Cluster Solution for Spinless Fermions}

Let us illustrate the essential steps that yield solutions to 
sign problems in the meron-cluster approach using the example of
spinless fermions. This is essentially a combination of results first 
discussed in \cite{Cha99,Cha00} and \cite{Cha00c}. This will serve 
as a preparation for the main subject of this paper --- namely fermions 
with spin. 

\subsection{Fermion Path Integral in Fock State Basis}

We consider fermions hopping on a $d$-dimensional cubic lattice of volume
$V = L^d$ sites with periodic or anti-periodic spatial boundary
conditions. The fermions are described by creation and annihilation
operators $c_x^\dagger$ and $c_x$ with standard anti-commutation relations
\begin{equation}
\{c_x^\dagger,c_y^\dagger\} = 0, \, \{c_x,c_y\} = 0, \, 
\{c_x^\dagger,c_y\} = \delta_{xy}.
\end{equation}
Following Jordan and Wigner \cite{Jor28} we represent these operators 
by Pauli matrices
\begin{equation}
c_x^\dagger = \sigma_1^3 \sigma_2^3 ... \sigma_{l-1}^3 \sigma_l^+, \,
c_x = \sigma_1^3 \sigma_2^3 ... \sigma_{l-1}^3 \sigma_l^-, \,
n_x = c_x^\dagger c_x = \frac{1}{2}(\sigma_l^3 + 1),
\end{equation}
where
\begin{equation}
\sigma_l^\pm = \frac{1}{2} (\sigma_l^1 \pm i \sigma_l^2), \, 
[\sigma_l^i,\sigma_m^j] = i \delta_{lm} \epsilon_{ijk} \sigma_l^k.
\end{equation}
Here $l$ labels the lattice point $x$. The Jordan-Wigner representation
requires an ordering of the lattice points. For example, one can label the 
point $x = (x_1,x_2,...,x_d)$ by $l = x_1 + x_2 L + ... + x_d L^{d-1}$.
It can be shown that the physics is completely independent of the 
ordering. Further, the Jordan-Wigner representation works in any dimension. 

Instead of considering the most general model that can be solved using the 
meron-concept, we restrict ourselves to a class of Hamiltonians 
that conserve particle number. From the discussion below it will become clear 
how to extend the ideas presented here to other Hamiltonians while maintaining 
the ability to solve the associated sign problem. With this in mind we
focus on a Hamilton operator of the form 
\begin{equation}
H = \sum_{x,i} h_{x,i},
\end{equation}
which is a sum of nearest-neighbor Hamiltonians $h_{x,i}$ given by
\begin{equation}
h_{x,i} = -\frac{t}{2}(c_x^\dagger c_{x+\hat i} + c_{x+\hat i}^\dagger c_x) + 
U (n_x - \frac{1}{2})(n_{x+\hat i} - \frac{1}{2}).
\end{equation}
which couples the fermion operators at the lattice sites $x$ and $x+\hat i$, 
where $\hat i$ is a unit-vector in the $i$-direction. Next we decompose the 
Hamilton operator into $2d$ terms
\begin{equation}
H = H_1 + H_2 + ... + H_{2d},
\end{equation}
with
\begin{equation}
H_i = \!\! \sum_{\stackrel{x = (x_1,x_2,...,x_d)}{x_i \rm{even}}} \!\! h_{x,i},
H_{i+d} = \!\! \sum_{\stackrel{x = (x_1,x_2,...,x_d)}{x_i \rm{odd}}} \!\! 
h_{x,i}.
\end{equation}
Note that the individual contributions to a given $H_i$ commute with each 
other, but two different $H_i$ do not commute. Using the Trotter-Suzuki formula
we express the fermionic grand canonical partition function as
\begin{eqnarray}
Z_f\!\!\!&=&\!\!\!\mbox{Tr} \{\exp[- \beta (H - \mu N)]\} \nonumber \\
&=&\!\!\!\lim_{M \rightarrow \infty} \! \mbox{Tr} 
\left\{\exp[- \epsilon (H_1 - \frac{\mu}{2d} N)] 
\exp[- \epsilon (H_2 - \frac{\mu}{2d} N)] ...
\exp[- \epsilon (H_{2d} - \frac{\mu}{2d} N)]\right\}^M. \nonumber \\ \,
\end{eqnarray}
Here $N = \sum_x n_x$ is the particle number operator and $\mu$ is the chemical
potential. We have introduced $M$ Euclidean time-slices with $\epsilon = 
\beta/M$ being the lattice spacing in the Euclidean time-direction. We insert 
complete sets of fermion Fock states between the factors 
$\exp[- \epsilon (H_i - \frac{\mu}{2d} N)]$. Each site is either empty or 
occupied, i.e.~$n_x$ has eigenvalue $0$ or $1$. In the Jordan-Wigner
representation this corresponds to eigenstates $|0\rangle$ and $|1\rangle$ of
$\sigma_l^3$ with $\sigma_l^3 |0\rangle = - |0\rangle$ and 
$\sigma_l^3 |1\rangle = |1\rangle$. The transfer matrix is a product of factors
\begin{eqnarray}
\label{transfer}
\lefteqn{\exp[- \epsilon (h_{x,i} - \frac{\mu}{2d} (n_x + n_{x+\hat i}))] = 
\exp[\epsilon(\frac{U}{4} + \frac{\mu}{2d})]}
\nonumber \\
&&\times
\left(\begin{array}{cccc} 
\exp[- \epsilon(\frac{U}{2} + \frac{\mu}{2d})] & 0 & 0 & 0 \\ 
0 & \cosh(\frac{\epsilon t}{2}) & \Sigma \sinh(\frac{\epsilon t}{2}) & 0 \\ 
0 & \Sigma \sinh(\frac{\epsilon t}{2}) & \cosh(\frac{\epsilon t}{2}) & 0 \\ 
0 & 0 & 0 & \exp[- \epsilon(\frac{U}{2} - \frac{\mu}{2d})] \end{array} \right),
\end{eqnarray}
which is a $4 \times 4$ matrix in the Fock space basis $|00\rangle$,
$|01\rangle$, $|10\rangle$ and $|11\rangle$ of two sites $x$ and $x+\hat i$.
Here $\Sigma = \sigma_{l+1}^3 \sigma_{l+2}^3 ... \sigma_{m-1}^3$ is a string 
of Pauli matrices running over consecutive labels between $l$ and $m$, where 
$l$ is the smaller of the labels for the lattice points $x$ and $x+\hat i$
and $m$ is the larger label. Note that $\Sigma$ 
is diagonal in the occupation number basis.

The partition function is now expressed as a path integral
\begin{equation}
Z_f = \sum_{[n]} \Sign[n] \exp(- S[n]),
\end{equation}
over configurations of occupation numbers $n(x,t) = 0,1$ on a 
$(d+1)$-dimensional space-time lattice of points $(x,t)$. The magnitude of
the Boltzmann factor, 
\begin{eqnarray}
&&\exp(- S[n]) = \nonumber \\
&&\prod_{p=0}^{M-1} \prod_{i=1}^{d} \prod_{\stackrel{x = (x_1,x_2,...,x_d)}
{x_i \rm{even},~t = 2dp+i-1}} \!\!\!\!\!\!\!\!
\exp\{- s[n(x,t),n(x+\hat i,t),n(x,t+1),n(x+\hat i,t+1)]\} \nonumber \\
&&\times\!\! \prod_{\stackrel{x = (x_1,x_2,...,x_d)}
{x_i \rm{odd},~t = 2dp+d+i-1}} \!\!\!\!\!\!\!\!\!
\exp\{- s[n(x,t),n(x+\hat i,t),n(x,t+1),n(x+\hat i,t+1)]\},
\end{eqnarray}
is a product of space-time plaquette contributions with
\begin{figure}
\begin{center}
\includegraphics[width=0.93\textwidth]{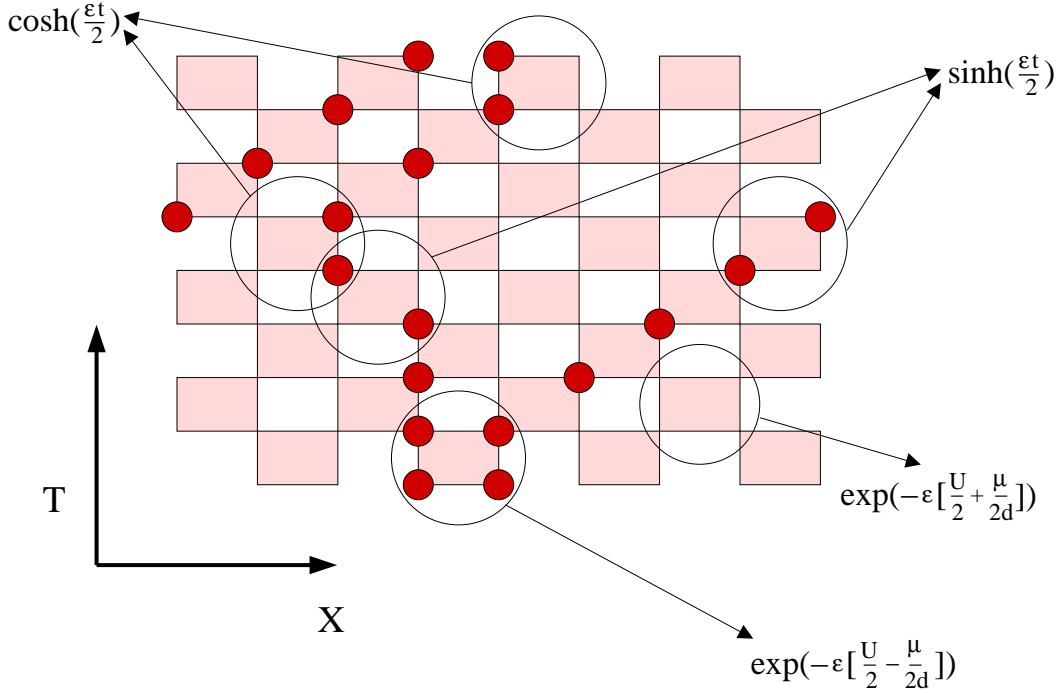}
\end{center}
\caption[]{\it A typical fermion world-line configuration in one spatial
dimension with periodic boundary conditions. The product of $\Sigma$
factors from each plaquette is
negative for this configuration since the two fermions exchange their
positions with each other.}
\label{fwlconf}
\end{figure}
\begin{eqnarray}
\label{Boltzmann}
&&\exp(- s[0,0,0,0]) = \exp[- \epsilon(\frac{U}{2} + \frac{\mu}{2d})], 
\nonumber \\
&&\exp(- s[0,1,0,1]) = \exp(- s[1,0,1,0]) = \cosh(\frac{\epsilon t}{2}),
\nonumber \\
&&\exp(- s[0,1,1,0]) = \exp(- s[1,0,0,1]) = \sinh(\frac{\epsilon t}{2}),
\nonumber \\
&&\exp(- s[1,1,1,1]) = \exp[- \epsilon(\frac{U}{2} - \frac{\mu}{2d})].
\end{eqnarray}
All the other Boltzmann factors are zero, which implies several constraints on
allowed configurations. Note that we have dropped the trivial overall factor 
$\exp[\epsilon(\frac{U}{4} + \frac{\mu}{2d})]$ in eq.(\ref{transfer}).
The sign of the Boltzmann factor $\Sign[n]$ also is a product of space-time 
plaquette contributions
$\mbox{sign}[n(x,t),n(x+\hat i,t),n(x,t+1),n(x+\hat i,t+1)]$ with
\begin{eqnarray}
&&\mbox{sign}[0,0,0,0] = \mbox{sign}[0,1,0,1] = \mbox{sign}[1,0,1,0] = 
\mbox{sign}[1,1,1,1] = 1, \nonumber \\
&&\mbox{sign}[0,1,1,0] = \mbox{sign}[1,0,0,1] = \Sigma.
\end{eqnarray}
It should be noted that the non-local string of Pauli matrices $\Sigma$ gets
contributions from all lattice points with labels between $l$ and $m$. This
would make an evaluation of the fermion sign rather tedious. Also, it is not
a priori obvious that $\Sign[n]$ is independent of the arbitrarily chosen order
of the lattice points. Fortunately, there is a simple way to compute 
$\Sign[n]$ which is directly related to the Pauli principle and which is
manifestly order-independent. In fact, $\Sign[n]$ has a topological meaning.
The occupied lattice sites define fermion world-lines which are closed around 
the Euclidean time-direction. Of course, during their Euclidean time-evolution
fermions can interchange their positions, and the fermion world-lines define a
permutation of particles. The Pauli principle dictates that the fermion sign is
just the sign of that permutation. When one works with anti-periodic spatial 
boundary conditions, $\Sign[n]$ receives an extra minus-sign for every fermion 
world-line that crosses the boundary. Figure \ref{fwlconf} shows a typical 
configuration in one spatial dimension along with examples of contributions 
to the Boltzmann weight arising from local space-time plaquettes. 

\subsection{Observables in Fermionic Fock Space}

A variety of observables of physical interest take on a simple form
in Fock space. Typically the expectation value of an operator
$O$ is given by
\begin{equation}
\langle O \rangle_f = \frac{1}{Z_f} \sum_{[n]} O[n] \Sign[n] \exp(- S[n]).
\end{equation}
For example, the total particle number relative to half-filling
\begin{equation}
N[n] - \frac{V}{2} = \sum_x \left(n(x,t) - \frac{1}{2}\right),
\end{equation}
which is conserved and hence the same in each time-slice, and its 
disconnected susceptibility 
\begin{equation}
\chi_N = \frac{\beta}{V} \left\langle 
\left(N - \frac{V}{2}\right)^2 \right\rangle_f
\end{equation}
are both operators that are diagonal in the occupation number basis.
Other quantities of physical interest are the staggered occupation
\begin{equation}
O[n] = \epsilon \sum_{x,t} (-1)^{x_1 + x_2 + ... + x_d}
 \left( n(x,t) - \frac{1}{2} \right),
\end{equation}
and the corresponding susceptibility
\begin{equation}
\chi_O = \frac{1}{\beta V} \left(\langle O^2 \rangle_f -
\langle O \rangle_f^2 \right).
\end{equation}
We can also measure the winding number $W_i[n]$ around the spatial
direction $i$
\begin{eqnarray}
W_i[n] &=& \! \sum_{p=0}^{M-1} \!\! \sum_{\stackrel{x = (x_1,x_2,...,x_d)}
 {x_i \rm{even}, ~ t = 2dp+i-1}} \!\!\!\!\!\!
 \frac{1}{L} \left\{
 \delta_{[n(x,t),n(x+\hat i,t),n(x,t+1),n(x+\hat i,t+1)],[1,0,0,1]}
 \right.
\nonumber \\
 && ~~~~~~~~~~~~~~~~~~~~~~ \left. - ~
 \delta_{[n(x,t),n(x+\hat i,t),n(x,t+1),n(x+\hat i,t+1)],[0,1,1,0]} \right\}
\nonumber \\
 &+& ~~~ \sum_{\stackrel{x = (x_1,x_2,...,x_d)}
 {x_i \rm{odd}, ~ t = 2dp+d+i-1}} \!\!\!\!\!\!\!\!
 \frac{1}{L} \left\{
 \delta_{[n(x,t),n(x+\hat i,t),n(x,t+1),n(x+\hat i,t+1)],[1,0,0,1]}
 \right.
\nonumber \\
 && ~~~~~~~~~~~~~~~~~~~~~~ \left. - ~
 \delta_{[n(x,t),n(x+\hat i,t),n(x,t+1),n(x+\hat i,t+1)],[0,1,1,0]}
 \right\}.
\end{eqnarray}
This counts $1$ for a plaquette configuration $[1,0,0,1]$, i.e., for a
fermion hopping in the positive $i$-direction, and $-1$ for
$[0,1,1,0]$, i.e., for a fermion hopping in the opposite direction. As
a result, $W_i[n]$ counts how many fermions wrap around the
$i$-direction during their Euclidean time-evolution.  Again, one can
define a corresponding susceptibility
\begin{equation}
\chi_{W_i} = \beta \langle W_i^2 \rangle_f.
\end{equation}
We will use a similar quantity as an order parameter for
superconductivity in models of fermions with spin. It measures the
response of the system to a twist in the spatial boundary
conditions. In the infinite volume limit the system feels the spatial
boundary only in a phase with long-range correlations.

Observables like the two-point Green function defined by
\begin{equation}
G(x,t;y,t') = \frac{
\mbox{Tr}\left\{\exp\left[-(\beta-t) H \right] c_x
\exp\left[-(t-t') H \right] c_y^\dagger \exp\left[-t' H \right]\right\}}
{\mbox{Tr}\left\{\exp\left[-\beta H \right]\right\}}
\end{equation}
are quite useful to extract the spectral information of $H$. These
expectation values of non-diagonal operators can also be obtained 
like those of the diagonal operators using
\begin{equation}
G(x,t;y,t') = \frac{1}{Z_f} \sum_{[n']} \Sign[n'] 
\exp(- S[n']).
\end{equation}
where the configurations $[n']$ that contribute are different from 
the configurations that contribute to the partition function due to the 
violation of fermion number at the space-time sites $(x,t)$ and $(y,t')$. 
In order to allow for these violations it is convenient to introduce two 
new time-slices at $t$ and $t'$. The factors $\exp(- S[n'])$ 
and $\Sign[n']$ can be calculated in the same way as before
on all the time-slices except on the two slices where fermionic 
creation and annihilation operators are introduced. At those two 
slices the fermion occupation number changes appropriately and
the Jordan-Wigner representation can be used to figure out the
extra sign factor coming from the string of $\sigma_3$ associated
with fermionic creation and annihilation operators. Interestingly,
as discussed in \cite{Bro98}, once the path integral
is re-written in terms of cluster variables, it is straightforward
to keep track of these new configurations along with the
configurations that contribute to the path integral and hence
such observables can also be measured.

\subsection{Cluster Decomposition of the Partition Function}

Up to now we have derived a path integral representation for the fermion
system in terms of bosonic occupation numbers and a fermion sign factor 
that encodes Fermi statistics. The system without the sign factor is bosonic 
and is characterized by the positive Boltzmann factor $\exp(- S[n])$. Here 
the bosonic model turns out to be a quantum spin system with the Hamiltonian
\begin{equation}
H = \sum_{x,i} [t (S_x^1 S_{x+\hat i}^1 + S_x^2 S_{x+\hat i}^2) + 
U S_x^3 S_{x+\hat i}^3] - \mu \sum_x S_x^3, 
\end{equation}
where $S_x^i = \frac{1}{2} \sigma_l^i$ is a spin $1/2$ operator associated with
the lattice site $x$ that was labeled by $l$. The case $U = t$ corresponds to
the isotropic anti-ferromagnetic quantum Heisenberg model, while $U = 0$ 
represents the quantum XY-model. The chemical potential plays the role of an 
external magnetic field.

It is well known that the partition function for the above quantum spin 
model can be re-written in terms of cluster variables. This cluster 
representation is at the heart of the the loop-cluster algorithm 
\cite{Eve93,Eve97} that is used to simulate the quantum spin
model very efficiently. Interestingly, the meron-cluster algorithm for 
the fermionic model is also based on a similar cluster decomposition of
the partition function. The essential differences come from the fermion 
sign factor and are encoded in the topology of the clusters. Once
the connection between the fermion permutation sign and the topology of 
the clusters is understood, the cluster algorithm for the quantum spin
system can be modified easily to deal with the fermionic model. 

\begin{figure}
\begin{center}
\includegraphics[width=0.93\textwidth]{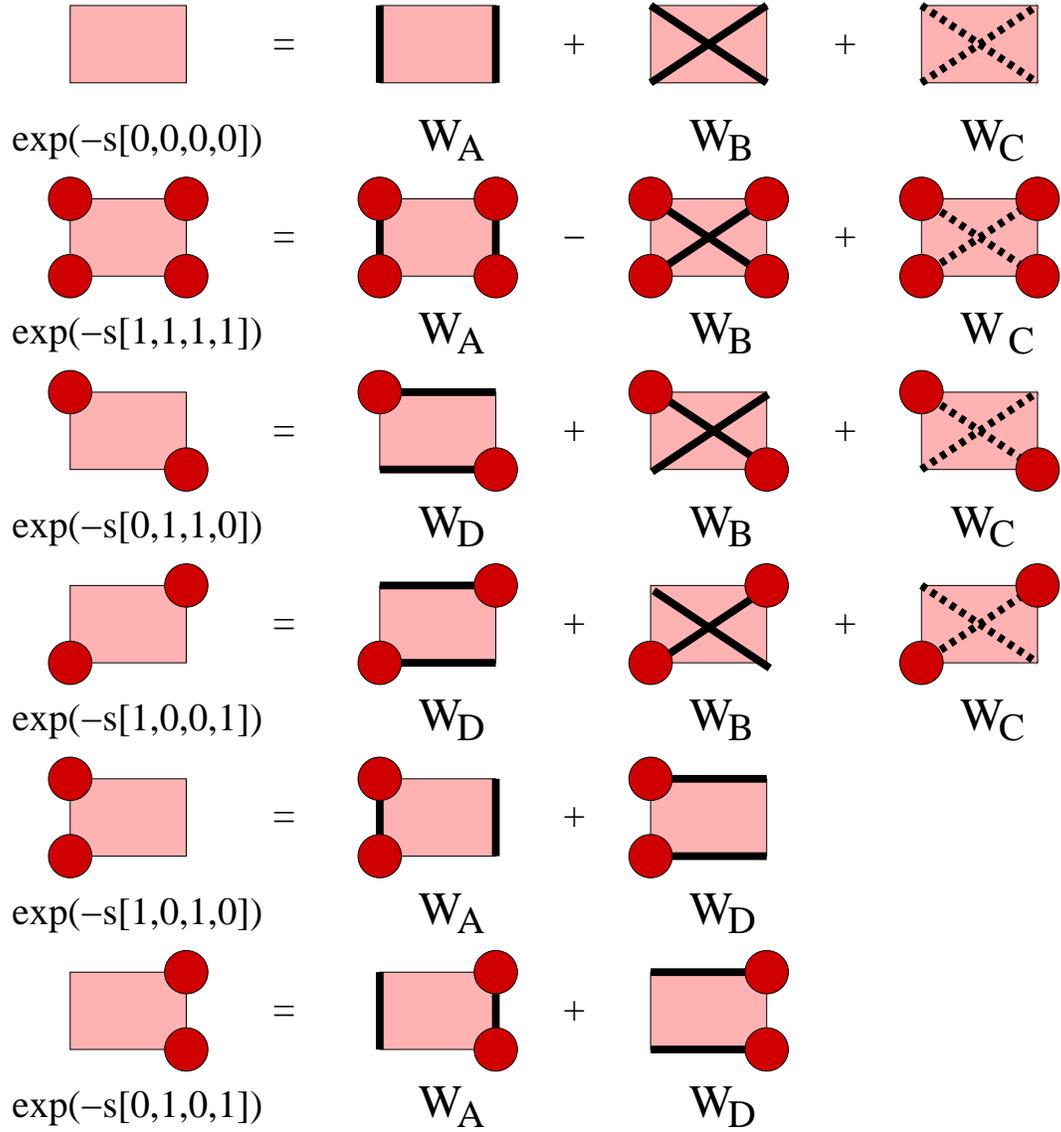}
\end{center}
\caption[]{\it The pictorial representation of the decomposition of 
plaquette weights of fermion occupation number configurations 
in terms of weights of bonds and fermion occupation numbers.}
\label{breakup}
\end{figure}

Let us first discuss the essential steps in the cluster decomposition 
of the partition function of the quantum spin model. The main idea 
is to take a configuration of quantum spins and re-write the weight as 
a sum over weights of configurations with spins and bonds, such that
the magnitude of the weight of the configuration does not change
when the set of spins connected by bonds are flipped. For the problem
at hand this decomposition can be accomplished at every plaquette. For 
example, the Boltzmann weights of each plaquette given in 
eq.(\ref{Boltzmann}) can be written as
\begin{eqnarray}
\label{decomp}
\lefteqn{\exp(- s[n(x,t),n(x+\hat i,t),n(x,t+1),n(x+\hat i,t+1)]) = }
\nonumber \\
&&\sum_{b=A,B,C,D}
\exp\left(-s[n(x,t),n(x+\hat i,t),n(x,t+1),n(x+\hat i,t+1);b]\right), 
\end{eqnarray}
where $b=A, B, C, D$ represent four types of bond break-ups. The
Boltzmann weights $\exp\left(-s[n;b]\right)$ can be arbitrary as
long as they satisfy eq.(\ref{decomp}). Here we choose
\begin{eqnarray}
\label{SBBoltzmann}
&&\exp\left(-s[0,0,0,0;A]\right)\;=\;
\exp\left(-s[1,0,1,0;A]\right)\;=\; 
\exp\left(-s[0,1,0,1;A]\right)\;=\;
\nonumber \\
&& \;\;\;\;\;\;\exp\left(-s[1,1,1,1;A]\right)\;=\;W_A\;=\;
\frac{1}{2}\left\{
\exp\left(-\epsilon \left[\frac{U}{2} + \frac{\mu}{2d}\right]\right)
+\exp\left(- \frac{\epsilon t}{2}\right) \right\},
\nonumber \\
&&\exp\left(-s[0,0,0,0;B]\right)\;=\;
\exp\left(-s[1,0,0,1;B]\right)\;=\;
\exp\left(-s[0,1,1,0;B]\right)\;=\;
\nonumber \\
&& \;\;\;\;\;\;-\exp\left(-s[1,1,1,1;B]\right)\;=\; W_B\;=\;
- \exp\left(- \frac{\epsilon U}{2} \right)\;
\sinh\left(\frac{\epsilon \mu}{2d}\right),
\nonumber \\
&&\exp\left(-s[0,0,0,0;C]\right)\;=\;
\exp\left(-s[1,0,0,1;C]\right)\;=\;
\exp\left(-s[0,1,1,0;C]\right)\;=\;
\nonumber \\
&& \;\;\;\;\;\;\exp\left(-s[1,1,1,1;C]\right)\;=\; W_C\;=\;
\frac{1}{2}\left\{
\exp\left(-\epsilon \left[\frac{U}{2} - \frac{\mu}{2d}\right]\right)
- \exp\left(- \frac{\epsilon t}{2}\right) \right\},
\nonumber \\
&&\exp\left(-s[1,0,1,0;D]\right)\;=\;
\exp\left(-s[0,1,0,1;D]\right)\;=\;
\exp\left(-s[0,1,1,0;D]\right)\;=\;
\nonumber \\
&& \;\;\;\;\;\;\exp\left(-s[1,0,0,1;D]\right)\;=\; W_D\;=\;
 \frac{1}{2}\left\{
\exp\left(\frac{\epsilon t}{2}\right)
- \exp\left(-\epsilon \left[\frac{U}{2} + \frac{\mu}{2d}\right]\right)
\right\}.\nonumber \\
\end{eqnarray}
to be the only non-zero weights. The above equations are shown
pictorially in figure \ref{breakup}.
If we interpret the spin states as fermion occupation numbers, clearly
$W[n,b]$ satisfies the first property discussed in section \ref{mcm}. 
We will now assume that $\mu \le 0$ and $t+\mu/d \geq U$,
so that all the Boltzmann 
weights $\exp(-s[n;b])$ of spin and bond configurations are 
positive. However, notice that there is an extra negative sign for each 
plaquette with configuration $[1,1,1,1;B]$. Such negative signs must
be considered separately as a contribution to the sign of a given 
spin and bond configuration. Explicitly this is given by $(-1)^{N_B}$ 
where $N_B$ is the number of plaquettes of the type $[1,1,1,1;B]$. 
Generally, such extra sign factors must be avoided since they can lead 
to sign problems. However, since we anyway have to deal with the non-local
fermion signs, which we have suppressed until now, we will allow for such 
extra negative sign factors. As we will see below, these local sign factors 
can sometimes be helpful in canceling sign factors that arise due to 
fermion permutations.

\begin{figure}
\begin{center}
\includegraphics[width=0.93\textwidth]{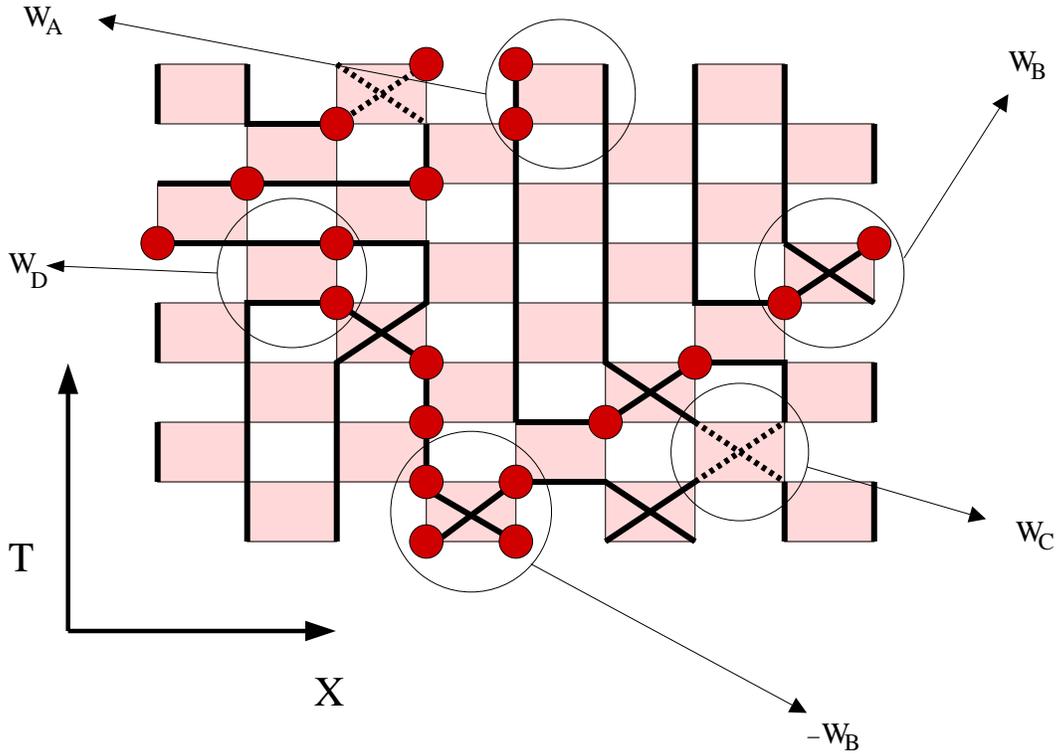}
\end{center}
\caption[]{\it A typical configuration of fermion occupation numbers and 
bonds in one spatial dimension.}
\label{sbconf}
\end{figure}

Based on the decomposition given in eq.(\ref{decomp}) the original
partition function of the fermionic model can be written as
\begin{equation}
\label{sbfpf}
Z_f = \sum_{[n,b]} \;\Sign[n,b]\; \exp(- S[n,b]),
\end{equation}
a sum over configurations of occupation numbers $n(x,t) = 0,1$ and
bond variables $b = A, B, C, D$ on each plaquette. The sign
$\Sign[n,b]$ is the product of the original fermion permutation
sign due to the occupation numbers and $(-1)^{N_B}$. The new
configurations naturally connect neighboring sites through the
bonds that live on each individual space-time interaction plaquette 
as represented pictorially in figure \ref{breakup}. A sequence of 
connected sites defines a cluster. Since each bond on a plaquette 
connects two sites, every cluster is a simple closed loop. Thus we
have written the partition function of a fermionic model in terms of the
statistical mechanics of closed interacting loops with not necessarily
positive
Boltzmann weights. Although we have illustrated this using a particular 
model, it is straightforward to repeat the above steps for any model.
When a cluster is flipped, the occupation numbers of all the sites 
on the cluster are changed from $n(x,t)$ to $1- n(x,t)$. In other 
words, a cluster-flip interchanges occupied and empty sites. The 
weights given in eq.(\ref{SBBoltzmann}) guarantee that the magnitude of
the weight of a configuration does not change when a cluster is flipped.
A typical configuration in one spatial dimension is represented 
schematically in figure \ref{sbconf}. 

\subsection{Meron-Clusters and Reference Configurations}

Let us now consider the effect of a cluster-flip on the sign
factor. In the example considered, the sign of a given configuration of
spins and bonds arises from two factors. One is the permutation of
fermion world-lines which is independent of the bond configuration and
the other is due to plaquettes of type $[1,1,1,1;B]$ each of which
contributes a negative sign to the Boltzmann weight. Obviously, a
cluster-flip either changes the sign of a configuration or not. In
general, the effect of the flip of a specific cluster on the fermion
sign depends on the orientation of the other clusters. For
example, a cluster whose flip does not change the sign now, may very
well change the sign after other clusters have been flipped. In other
words, in general the clusters affect each other in their effect on the
configuration's sign. Recently, a formula for how cluster-flips affect
the fermion permutation sign was discovered \cite{Cha00c}. The relevant 
information
is encoded by an integer which can be determined by starting at an
arbitrary point on the loop-cluster and traversing around it in either
direction and flipping the sites as they are encountered. As we go around 
the loop, we denote the number of horizontal bonds ($D$-type bonds)
encountered
while entering an occupied site from an empty site (before they are flipped)
by $N_h$, the number of cross-bonds ($B$- or $C$-type bonds) traversed
while the 
other cross-bond on the same plaquette connects two occupied sites by $N_x$, 
and the temporal winding of the cluster by $N_w$. Then the cluster-flip 
changes the fermion permutation sign if and only if
\begin{equation}
\label{flipeq}
N = N_h + N_w + N_x
\end{equation}
is even. For details of how this can be derived we refer the reader to 
\cite{Cha00c}. A similar formula can be derived for the signs 
associated with the plaquettes of the type $[1,1,1,1;B]$. In this case, again
as we traverse the loop-cluster and flip the sites, if $N_{xB}$ is the 
number of $B$-type cross-bonds encountered when the other cross-bond 
connects two occupied sites, then the cluster-flip changes the sign
associated with the plaquettes of the type $[1,1,1,1;B]$ if $N_{xB}$ is odd. 
Using this formula we discover that in the present model the effect of
a cluster-flip on the sign of a configuration in general 
depends on the orientation of other clusters. In particular, if two
clusters cross each other in an odd number of $C$-type cross-bonds then 
the flip of one cluster has an effect on whether the other 
cluster-flip changes the configuration sign or not. Interestingly, 
once we forbid $C$-type cluster break-ups, i.e.~when $W_C=0$, the clusters 
have a remarkable property with far reaching consequences: each cluster can 
be characterized by its effect on the fermion sign independent of the 
orientation  of all the other clusters. We refer to clusters whose flip 
changes $\Sign[n,b]$ as merons, while clusters whose flip leaves $\Sign[n,b]$ 
unchanged are called non-merons. The flip of a meron-cluster permutes the 
fermions and changes the topology of the fermion world-lines. Notice that 
the $B$-type cross-bonds are still allowed thanks to the extra negative 
signs that arise due to signs from $[1,1,1,1;B]$ plaquettes. Thus we have
been able to ensure the first two properties discussed in section \ref{mcm},
provided we satisfy the constraints $\mu \leq 0$, $t + \mu/d \geq U$ and 
$W_C = 0$ which implies $t= U - \mu/d$.

When the cluster-flips affect the sign factor independently, it is easy to
average analytically over all the cluster-flips. 
Such an average over cluster-flips for
a given bond configuration is referred to as an improved estimator in the
language of cluster algorithms. We will discuss improved estimators of
some observables in the next section. Averaging the sign over cluster-flips, 
the fermionic partition function can be re-written as 
\begin{equation}
Z_f = \sum_{[n,b]} \;\overline{\Sign[b]}\; \exp(- S[n,b]),
\end{equation}
where
\begin{equation}
\overline{\Sign[b]}\;=\;\frac{1}{2^{N_{\cal C}}}\;\sum_{\rm{cluster-flips}}
\Sign[n,b].
\end{equation}
Thus $\overline{\Sign[b]}=0$ if at least one of the clusters is a meron.
In order to solve the sign problem it is important to eliminate configurations
with meron-clusters from the Monte Carlo sampling. Unfortunately, this
alone does not guarantee that the 
contributions from the zero-meron sector will always be positive,
although it may alleviate the sign problem considerably. With no merons 
in the configuration it is clear that the sign of a configuration
remains unchanged under cluster-flips, but one could still have 
$\Sign[n,b] = - 1$. In order to solve the sign problem completely
it is necessary to show that configurations with no merons are always
positive. This takes us to the third property discussed in section
\ref{mcm}, the need for a reference configuration with a positive
weight. Interestingly, there exists a class of models where this is 
achievable. It is then easy to show that $\overline{\Sign[b]}=1$ in 
the non-meron sector. This solves the sign problem completely. Thus
the fermionic partition function is obtained as a sum over weights 
of configurations in the zero-meron sector,
\begin{equation}
Z_f \;=\; \sum_{[b],{\rm zero-meron}} 2^{N_c} W[b],
\label{clpf1}
\end{equation}
where $W[b]$ is obtained as a product of local weights $W_A$, $W_B$ and $W_D$.

There are at least two types of reference configurations that arise
for spinless fermions. For example, in addition to setting
$W_C=0$, if we eliminate the $B$-type bonds completely by also setting
$W_B=0$, the remaining cluster break-ups have a remarkable property. 
They guarantee that sites inside a cluster obey a pattern of staggered 
occupation, i.e.~either the even sites (with $x_1 + x_2 + ... + x_d$ even) 
within the cluster are all occupied and the odd sites are all empty or
vice versa. This guarantees that the clusters can be flipped such that 
one reaches the totally staggered configuration in which all even sites
at every time-slice are occupied and all odd sites are empty. 
In this half-filled configuration all fermions are static, no fermions are 
permuted during the Euclidean time-evolution. Further, since the model
has no other sign factors, we know that $\Sign[n_{\rm{ref}},b] = 1$. Since any 
configuration can be turned into the totally staggered configuration by 
appropriate cluster-flips, property three is indeed satisfied.
The condition $W_B=0$ and $W_C=0$ leads to the restriction $\mu=0$ and 
$t=U$. This leads to an interesting interacting spinless fermion model.
A variant of this model has been studied in \cite{Cha99,Cha00a,Cha00b,
Cox00} where it has been shown that the model can be used to study
the spontaneous breaking of a $\mathbf{Z}(2)$ symmetry.
\begin{figure}[ht]
\begin{center}
\vskip0.1in
\hbox{
\includegraphics[width=0.45\textwidth]{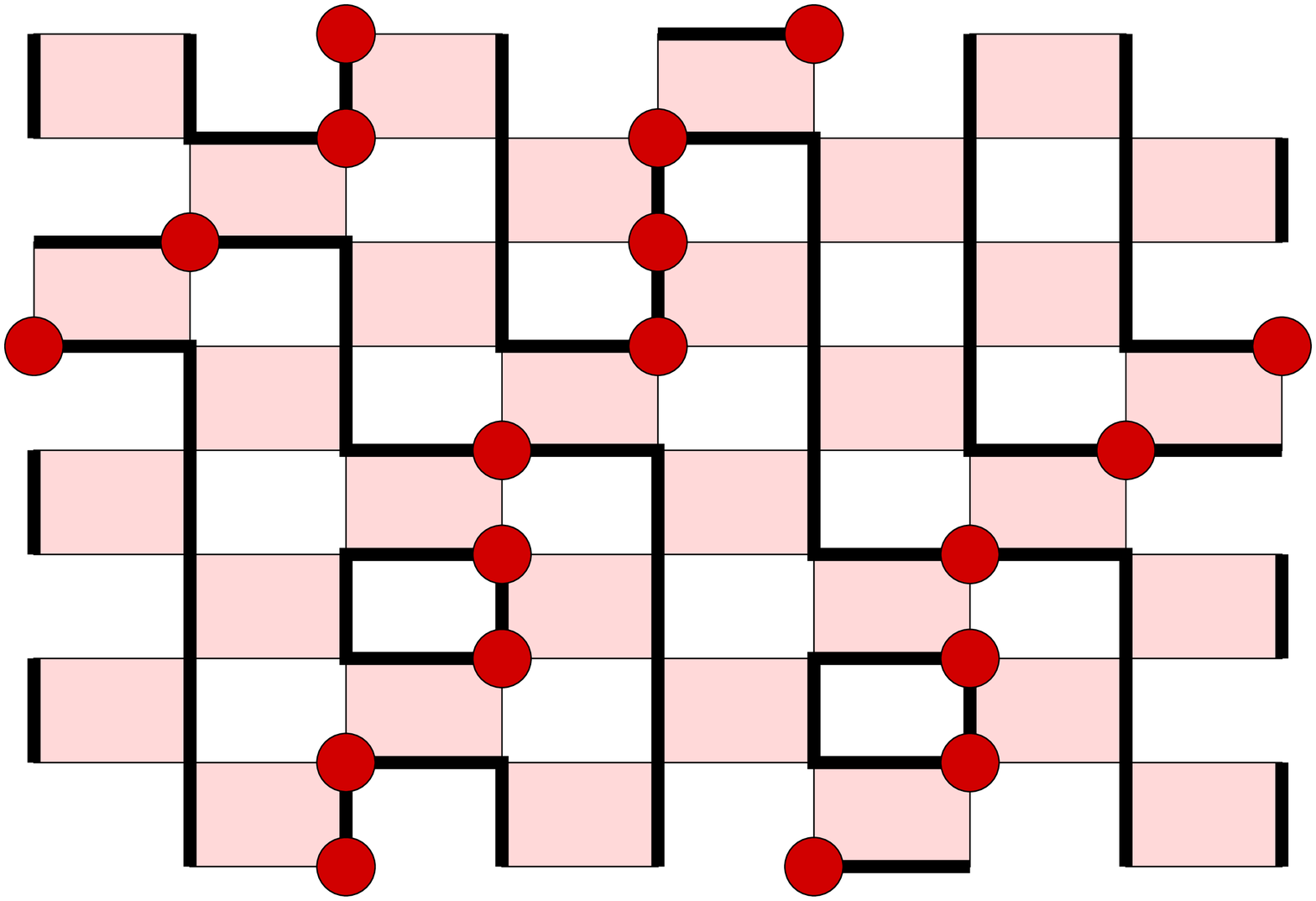}
\hskip0.4in
\includegraphics[width=0.45\textwidth]{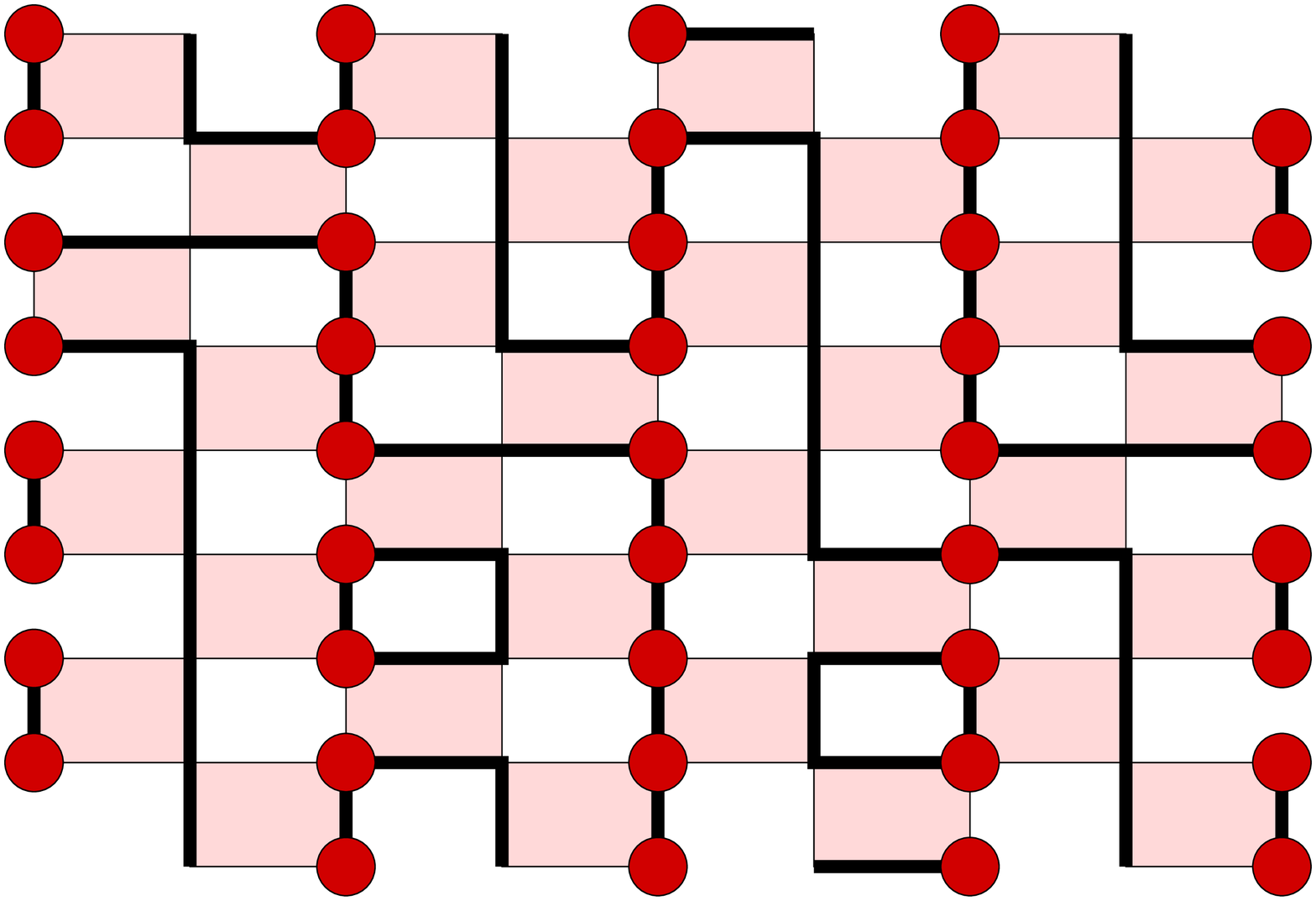}
}
\vskip0.1in
\hbox{
\hskip1.4in (a)
\hskip2.8in
(b)
}
\vskip0.2in
\hbox{
\includegraphics[width=0.45\textwidth]{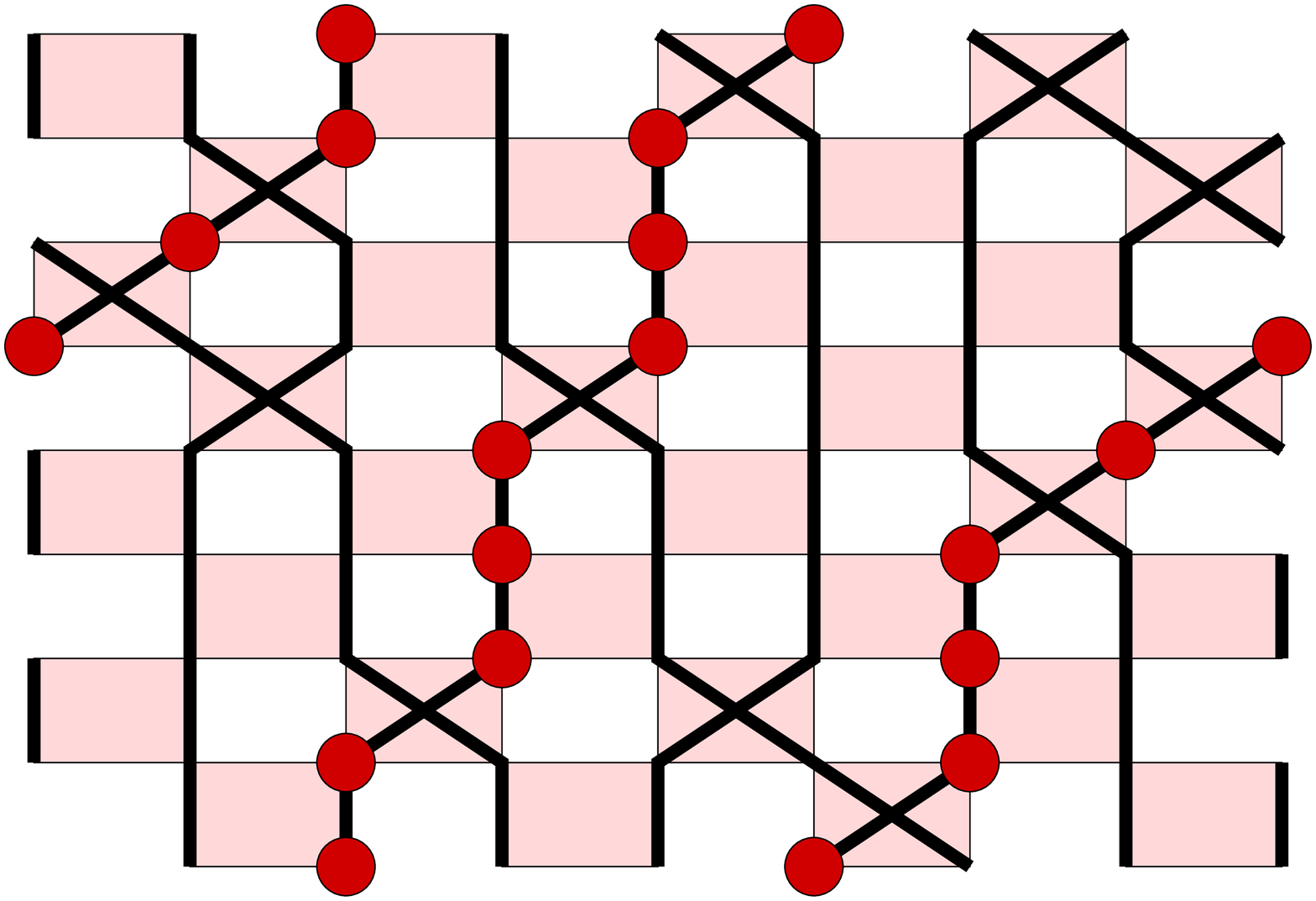}
\hskip0.4in
\includegraphics[width=0.45\textwidth]{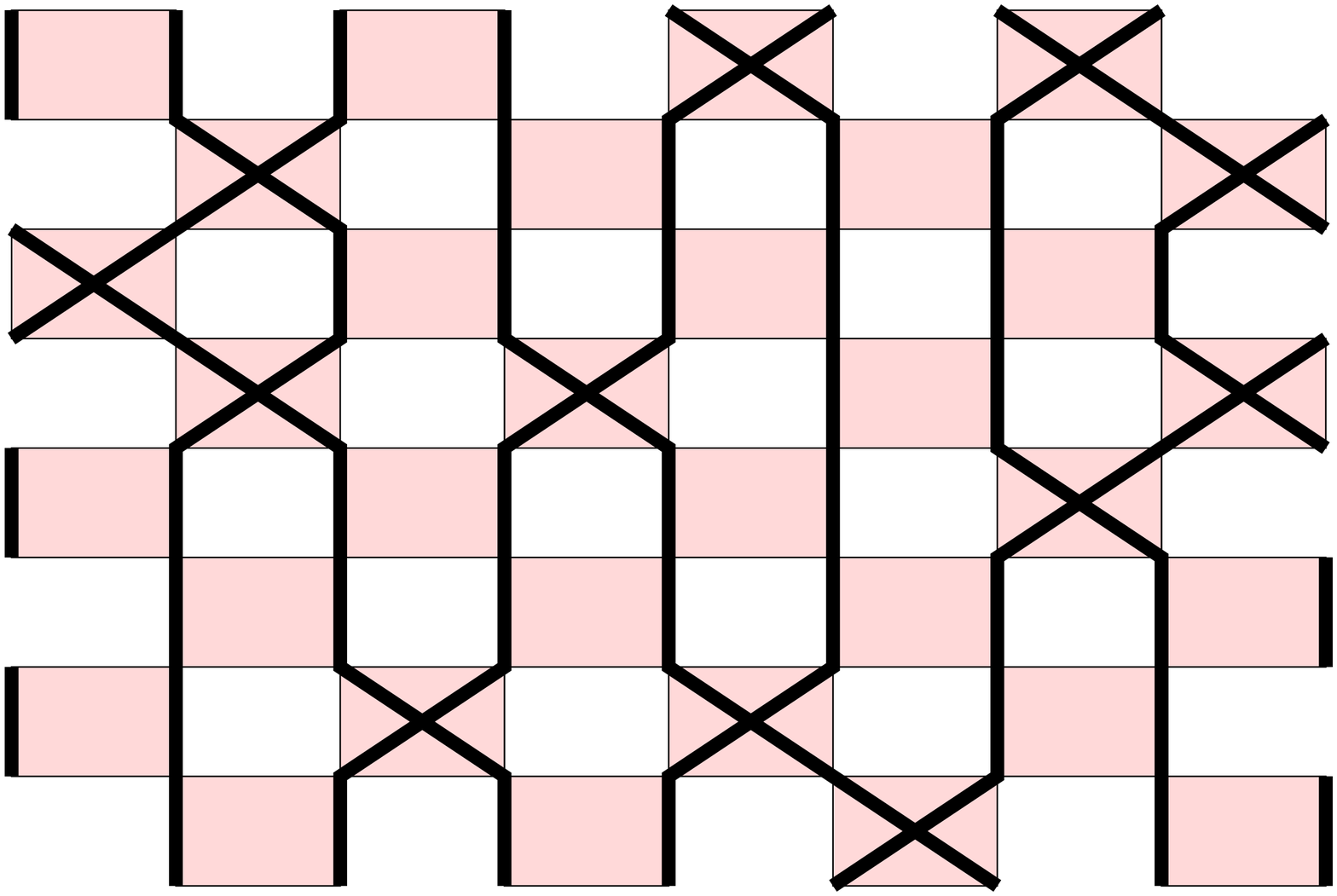}
}
\vskip0.1in
\hbox{
\hskip1.4in (c)
\hskip2.8in
(d)
}
\end{center}
\caption{\it Figure (a) shows a typical fermion and bond configuration
without cross-bonds and (b) shows the corresponding reference 
configuration which has a positive sign. Figures (c) and (d)
show similar configurations for the case without horizontal 
bonds.\label{refconfs}}
\end{figure}

The second reference configuration is obtained by setting 
$W_D=0$ along with $W_C=0$. In this case we see that all sites in
a given cluster have the same occupation number, i.e., all are either
empty or all are occupied. This means that all clusters can be flipped
such that one reaches a configuration where all sites are empty.
Since there are no fermions at all, there is no fermion permutation
sign in this configuration. Further, there are no negative signs
because of plaquettes of the form $[1,1,1,1;B]$, since all sites
are empty and $[0,0,0,0;B]$ is positive. Therefore the configurations
with no merons necessarily have a positive sign. 
The condition $W_C=0$ and $W_D=0$ leads to the restriction $U=0$ and 
$t=-\mu/d$ with $\mu \leq 0$. This model leads to free non-relativistic 
fermions and has been discussed in \cite{Cha00}. 

Figure \ref{refconfs} shows a typical configuration and the associated
reference configuration for each of the two classes discussed
above. The reference configurations are closely connected with
the ground state properties of each model. The staggered reference
configuration naturally describes theories with spontaneously broken
ground states whereas the totally empty configuration leads to free
non-relativistic fermions. It is possible to include a limited form of
additional interactions without destroying any of the useful
properties in these two classes of theories. However, it would be
interesting to find new reference configurations which can lead to
interesting physics. Of course, there are a variety of non-translationally
invariant reference configurations which could describe electronic
properties of certain crystalline structures. These models may be worth 
exploring. As we will show below, novel reference configurations also
arise when spin is introduced.

\subsection{Building-Blocks of Cluster Models}
\label{bldblks}

\begin{figure}[ht]
\begin{center}
\includegraphics[width=0.9\textwidth]{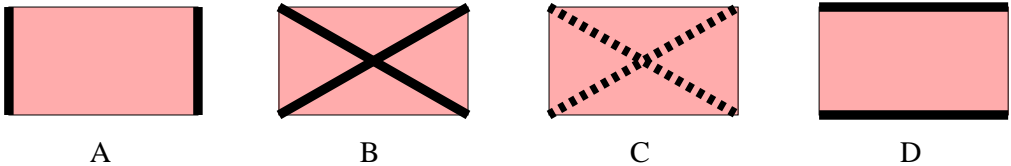}
\end{center}
\caption{\it Building-blocks of the two-site interactions discussed in
the text.}
\label{2sbldblks}
\end{figure}

The results of the last two sections can be used to synthesize
simple rules that help in building models of spinless fermions which
automatically satisfy the first two properties of the meron-cluster 
approach, i.e. the weight of a configuration of clusters does not 
change under a cluster-flip and the effect on the sign of a 
configuration due to a cluster-flip is independent of all other clusters. 
In the models constructed above we started with four types of bond 
interactions namely the $A$-, $B$-, $C$- and $D$-types as shown in the figure 
\ref{2sbldblks}. Each of these interactions can be associated with an 
operator acting on the two-site Hilbert space. For example, a plaquette 
with $A$-type (vertical) bonds gives the same positive 
weight to all diagonal elements. Hence $A$-type bonds are 
equivalent to the unit matrix. Similarly, the $B$-, $C$- and $D$-type bonds 
are equivalent to operators tabulated in table \ref{tab2sbldblks}. The 
fermionic nature of the creation and annihilation operators are encoded 
in the formula of eq.(\ref{flipeq}) which is used to figure
out the relative sign of configurations when clusters are flipped.
\begin{table}[ht]
\begin{center}
\begin{tabular}{|c|c|}
\hline
Bond-type & Transfer Matrix \\
\hline
$A$ & $\mathbf{1}$ \\
$B$ & $c_x^\dagger c_y + c_y^\dagger c_x - n_x - n_y + 1$ \\
$C$ & $c_x^\dagger c_y + c_y^\dagger c_x + 2(n_x - 1/2)(n_y-1/2) + 1/2$ \\
$D$ & $c_x^\dagger c_y + c_y^\dagger c_x - 2(n_x - 1/2)(n_y-1/2) + 1/2$ \\
\hline
\end{tabular}
\end{center}
\caption{\it \label{tab2sbldblks}
The two-site building-blocks of figure \ref{2sbldblks} and their 
transfer matrix contributions.}
\end{table}
Using eq.(\ref{flipeq}) along with the weights of the building-blocks,
we found that type $C$ bonds are forbidden as they violate the required
property that cluster-flips should have an independent effect on the sign.
Thus the $A$-, $B$- and $D$-type bonds provide the basic building-blocks 
for constructing Hamiltonians which satisfy the two necessary 
requirements of the meron-cluster approach. We only need to ensure the
existence of a reference configuration. This restriction then forbids 
models which simultaneously allow $B$- and $D$-type bonds. Thus, we
find two types of solvable models discussed above. It is easy to construct
the Hamiltonian of these models using the corresponding building
blocks. For example, 
the model with $A$- and $B$-type bonds leads to the two-site transfer matrix 
$T_{xy} = W_A +  W_B[c_x^\dagger c_y + c_y^\dagger c_x - n_x - n_y + 1]$, 
which in the limit $\epsilon\rightarrow 0$ yields $T_{xy} = 1 +  
\epsilon[c_x^\dagger c_y + c_y^\dagger c_x - n_x - n_y + 1]$.
Since all two-site interactions contribute equally to the full
Hamiltonian, $H=\sum_{<xy>} h_{xy}$, in this limit we can identify 
$T = 1 - \epsilon H$, which yields 
\begin{equation}
h_{xy} = -(c_x^\dagger c_y + c_y^\dagger c_x) + n_x + n_y
\label{freef}
\end{equation}
up to additive and multiplicative constants.
Similarly the model obtained with $A$- and $D$-type bonds leads to
\begin{equation}
h_{xy} = -(c_x^\dagger c_y + c_y^\dagger c_x) + 2(n_x-1/2)(n_y-1/2).
\label{slhubb}
\end{equation}
Thus, one can construct the Hamilton operator once the building-blocks
are chosen in terms of the bond break-ups on an elementary plaquette. 
Later, we will use this procedure to construct models of fermions with 
spin. In eq. (\ref{freef}) and eq. (\ref{slhubb}) we have chosen to
normalize the hopping term and set $t=1$. In the remainder of this
article we will continue to adopt this normalization for convinence.

\subsection{Improved Estimators}
\label{is1}

Based on the previous discussion it is clear that the partition 
function can be written entirely in terms of cluster variables as
\begin{equation}
Z_f \;=\; \sum_{[b],{\mathrm{zero-meron}}} 2^{N_C} ~ W[b]
\end{equation}
where $N_C$ is the total number of clusters and the sum is over cluster
configurations without meron-clusters. Similarly, it is possible to 
find expressions for observables in the language of clusters. For
example, consider the deviation of the average occupation number 
of a configuration from half-filling which is defined as
\begin{equation}
\Delta n \;=\; 
n - \frac{1}{2} = \frac{1}{V} \sum_x \left(n_x - \frac{1}{2}\right).
\end{equation}
Clearly it must be possible to sum over $x$ belonging to a 
particular cluster ${\cal C} \in [b]$ first and then sum over all 
possible clusters. If we define
\begin{equation}
\Omega_{{\cal C}} = \sum_{x\in{\cal C}} (n_x-1/2),
\end{equation}
it is easy to see that $\Delta n = (1/V) \sum_{\cal C} \Omega_{\cal C}$. 
Performing a partial average over flips of fermionic occupation numbers 
belonging to various clusters one finds that this average is non-zero 
only in the one-meron sector and in this sector
\begin{equation}
\left<\Delta n\right>_{\rm{cluster-flips}} \;=\;
\frac{1}{V}\;\Omega_{{\cal C}_{\rm meron}}
\end{equation}
where $\Omega_{{\cal C}_{\rm meron}}$ is to be calculated for the
meron-cluster which is flipped into the reference configuration. 
Thus the complete answer turns out to be
\begin{equation}
\langle \Delta n \rangle = \frac{1}{Z_f} \sum_{[b],\mathrm{one-meron}}
\frac{\Omega_{{\cal C}_{\rm{meron}}}}{V} 2^{N_C} W[b].
\end{equation}
In the case of spinless fermion models discussed above,
it is easy to see that $\Omega_{\cal C}$ is equal to the temporal 
winding of the cluster ${\cal C}$ up to a factor of 2. 
For the model described by the  Hamiltonian of eq.(\ref{slhubb}), 
$\langle \Delta n \rangle = 0$ since there is no contribution from
the single-meron sector to the partition function.
Similar improved estimators for the 
chiral condensate \cite{Cha00b} and susceptibility \cite{Cha00a} can be 
constructed. In section \ref{is2} we will consider improved estimators for 
the susceptibilities relevant for the study of superconductivity.

\section{Models of Fermions with Spin}

There are many ways to construct models of fermions with spin such
that the three properties necessary for the meron-cluster method
to work, as discussed in section \ref{mcm}, are satisfied. The most
general approach is to start from the Hamilton operator of interest
and proceed with the same steps as discussed in the earlier section.
If the Hamilton operator is a sum of only two-fermion interactions,
then nothing qualitatively new emerges since it is possible to consider 
the spin degree of freedom as two ``spatial'' layers representing up
and down spins. 
All on-site interactions that can occur between the two spin layers 
can be introduced on a separate time-slice. However, it is now possible 
to find new reference configurations which are physically meaningful and 
new models can be constructed using them. 

\begin{figure}
\begin{center}
\includegraphics[width=0.93\textwidth]{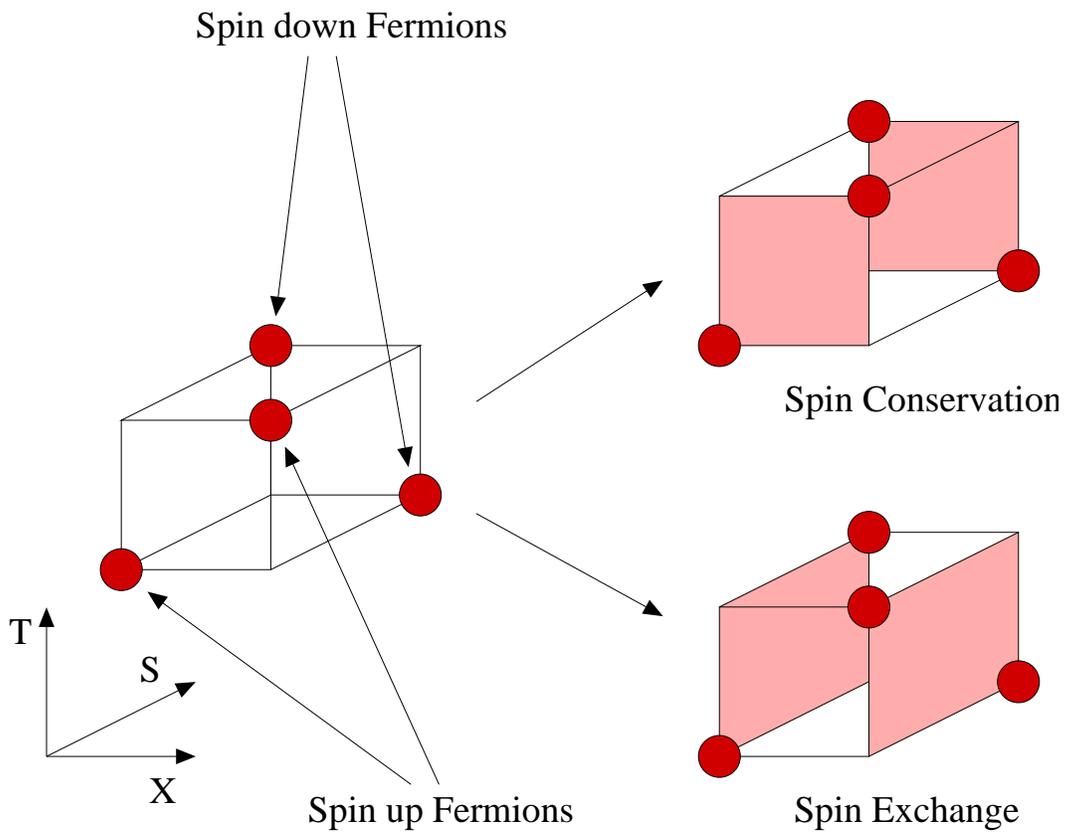}
\end{center}
\caption[]{\it Configurations with four-site 
interactions can arise as correlated two-site interactions in different 
ways. The two possible interactions result in different fermion permutation
signs.}
\label{4siteint}
\end{figure}
On the other hand, it is natural to allow four-``site'' interactions
which connect nearest spatial neighbors and the two spin layers. In this case
new situations not considered in the previous sections can arise.
As an example, consider the transfer matrix element given in figure 
\ref{4siteint}. Although such four-site interactions have not been 
considered earlier, they can be thought of as two correlated two-site 
interactions. For example, the shown interaction can arise either 
due to a ``correlated'' fermion hop which conserves spin or a 
``correlated'' spin-flip. These two alternatives are shown on the 
right-hand side in the same figure. If we assume that these new
interactions arise from such correlated two-site interactions, then 
it is possible to use the technology developed for spinless fermions. 
Here we will take this approach and consider models with four-site
interactions.

Instead of trying to find the most general solvable model, we 
again restrict ourselves to a class of models that can be built with
simple correlated two-site interactions. In the next section we will
enumerate some of the simplest correlated two-site interactions and
in the following section we will build one of the simplest
solvable models which has the symmetries of the Hubbard model.

\subsection{Building-Blocks of a Cluster Model}

\begin{figure}
\begin{center}
\includegraphics[width=0.93\textwidth]{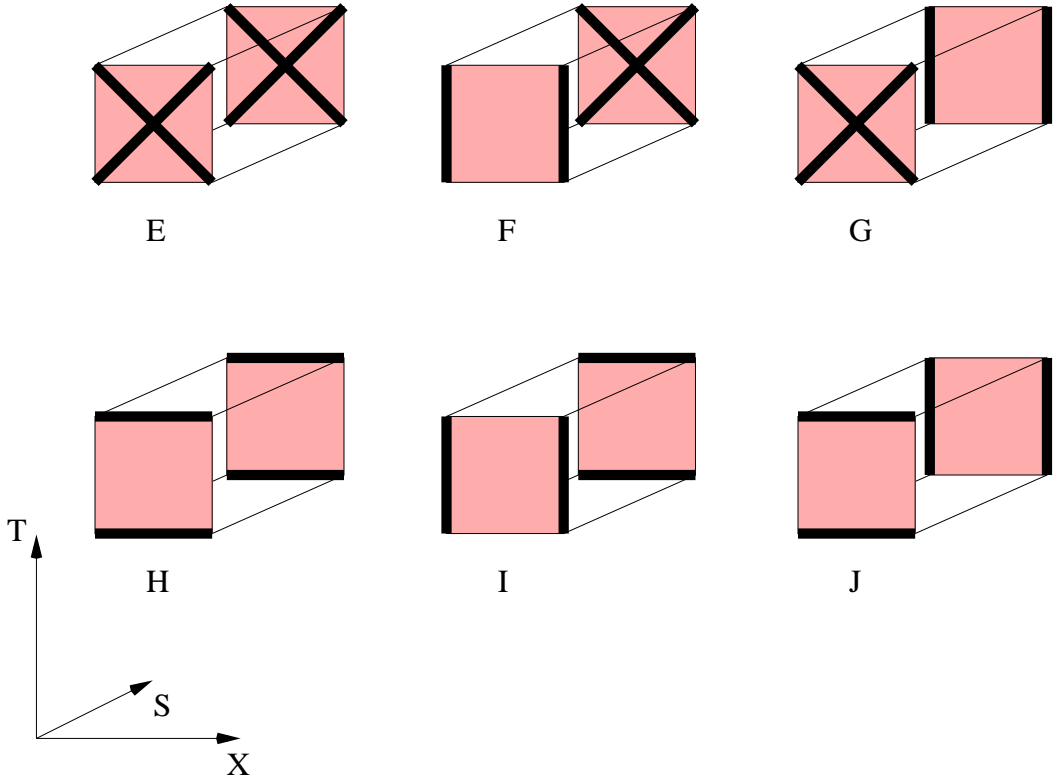}
\end{center}
\caption[]{\it Spin-conserving bond break-ups that form the building-blocks
of models discussed in the text.}
\label{4sbldblks}
\end{figure}

Let us first discuss the simplest building-blocks for
constructing models of fermions with spin that can be solved using
meron-cluster techniques. In the case of spinless
fermions, these building-blocks were discussed in section
\ref{bldblks}. There it was shown how these elementary blocks satisfy
two of the necessary properties and how one can build models with a
reference configuration that ensures that all the criteria of
solvability are met. We also discussed how it is possible to find the
Hamilton operator for a model that is solvable using the meron-cluster
approach. In the case of fermions with spin the various bond break-ups
$(E,F,G,H,I,J)$ which conserve spin are shown pictorially in figure
\ref{4sbldblks}.
In table \ref{tab4sbldblks} the transfer matrix elements associated with
each of the building-blocks is tabulated.
\begin{table}[hbt]
\begin{center}
\begin{tabular}{|c|c|}
\hline
Bond-type & Transfer Matrix \\
\hline
$E$ &  $[{\cxd}^\dagger \cyd + {\cyd}^\dagger \cxd - \nxd - \nyd + 1]$ \\
& $\times [{\cxu}^\dagger \cyu + {\cyu}^\dagger \cxu - \nxu - \nyu + 1]$ \\
$F$ & ${\cxd}^\dagger \cyd + {\cyd}^\dagger \cxd - \nxd - \nyd + 1$ \\
$G$ & ${\cxu}^\dagger \cyu + {\cyu}^\dagger \cxu - \nxu - \nyu + 1$ \\
$H$ & $[{\cxd}^\dagger \cyd + {\cyd}^\dagger \cxd - 
2(\nxd - 1/2)(\nyd - 1/2) + 1/2]$ \\ 
& $\times [{\cxu}^\dagger \cyu + {\cyu}^\dagger \cxu - 
2(\nxu - 1/2)(\nyu - 1/2) + 1/2]$ 
\\
$I$ & ${\cxd}^\dagger \cyd + {\cyd}^\dagger \cxd - 
2(\nxd - 1/2)(\nyd-1/2) + 1/2$ \\
$J$ & ${\cxu}^\dagger \cyu + {\cyu}^\dagger \cxu - 
2(\nxu - 1/2)(\nyu-1/2) + 1/2$ \\
\hline
\end{tabular}
\end{center}
\caption{\it The operators associated with the 
building-blocks of spin conserving Hamiltonians shown in figure 
\ref{4sbldblks}.}
\label{tab4sbldblks}
\end{table}
Using the building-blocks it is possible to find models which satisfy
all the three required properties of solvability discussed in section
\ref{mcm}. As an illustration we construct a Hubbard-type model below.

\subsection{A Hubbard-Type Model}

\label{hubbmodel}

The Hubbard model is one of the simplest models used to describe
superconductivity. In particular, it is believed that the 2-dimensional
repulsive Hubbard model may describe the physics of high-$T_c$ cuprate 
superconductors \cite{Sca00}. The Hamilton operator of 
the Hubbard model is given by
\begin{eqnarray}
H &=& \sum_{<xy>, s=\uparrow,\downarrow} 
\left\{ c_{x s}^\dagger c_{y s} + c_{y s}^\dagger c_{x s}\right\} 
\nonumber \\
&& \;\;\;\;\;\;\;\;+\; \sum_{x} 
\left\{U\left(\nxu-\frac{1}{2}\right)\left(\nxd-\frac{1}{2}\right)
\;-\;\mu (\nxu + \nxd)\right\}
\end{eqnarray}
This model has a global $SU(2)$ spin symmetry whose 
generators are given by
\begin{equation}
S^+ = \sum_x \cxu^\dagger \cxd,\;\;\;
S^- = \sum_x \cxd^\dagger \cxu,\;\;\;
S_3 = \frac{1}{2}\sum_x (\nxu - \nxd)
\label{sgen}
\end{equation}
and a global $SU(2)$ pseudo-spin symmetry at $\mu = 0$ whose generators 
in two dimensions are given by
\begin{equation}
 J^+ = \sum_x (-1)^{x_1+x_2} \cxu^\dagger \cxd^\dagger,\;\;\;
 J^- = \sum_x (-1)^{x_1+x_2} \cxd \cxu,\;\;\;
 J_3 = \frac{1}{2}\sum_x (\nxu + \nxd - 1)
\label{jgen}
\end{equation}
The chemical potential breaks the pseudo-spin symmetry to a
$U(1)$ fermion number symmetry. Superconductivity is a result of the
spontaneous breaking of this $U(1)$ symmetry. In two dimensions, due
to the Mermin-Wagner theorem, it is believed that this symmetry
breaking must follow the Kosterlitz-Thouless predictions \cite{KT72}.
For $U<0$ the theory is expected to show s-wave superconductivity at
low temperatures \cite{Sca89, Mor91}. Then the model can be formulated
such that there is no sign problem in traditional quantum
Monte Carlo algorithms. However, due to the difficulty of these
conventional fermion algorithms the determination of the critical
temperature is still controversial \cite{Lac98}. The repulsive model
with $U>0$, on the other hand, is more interesting but suffers from a
sign problem away from half-filling \cite{Whi89}.

\begin{figure}[ht]
\begin{center}
\hbox{
\includegraphics[width=0.48\textwidth]{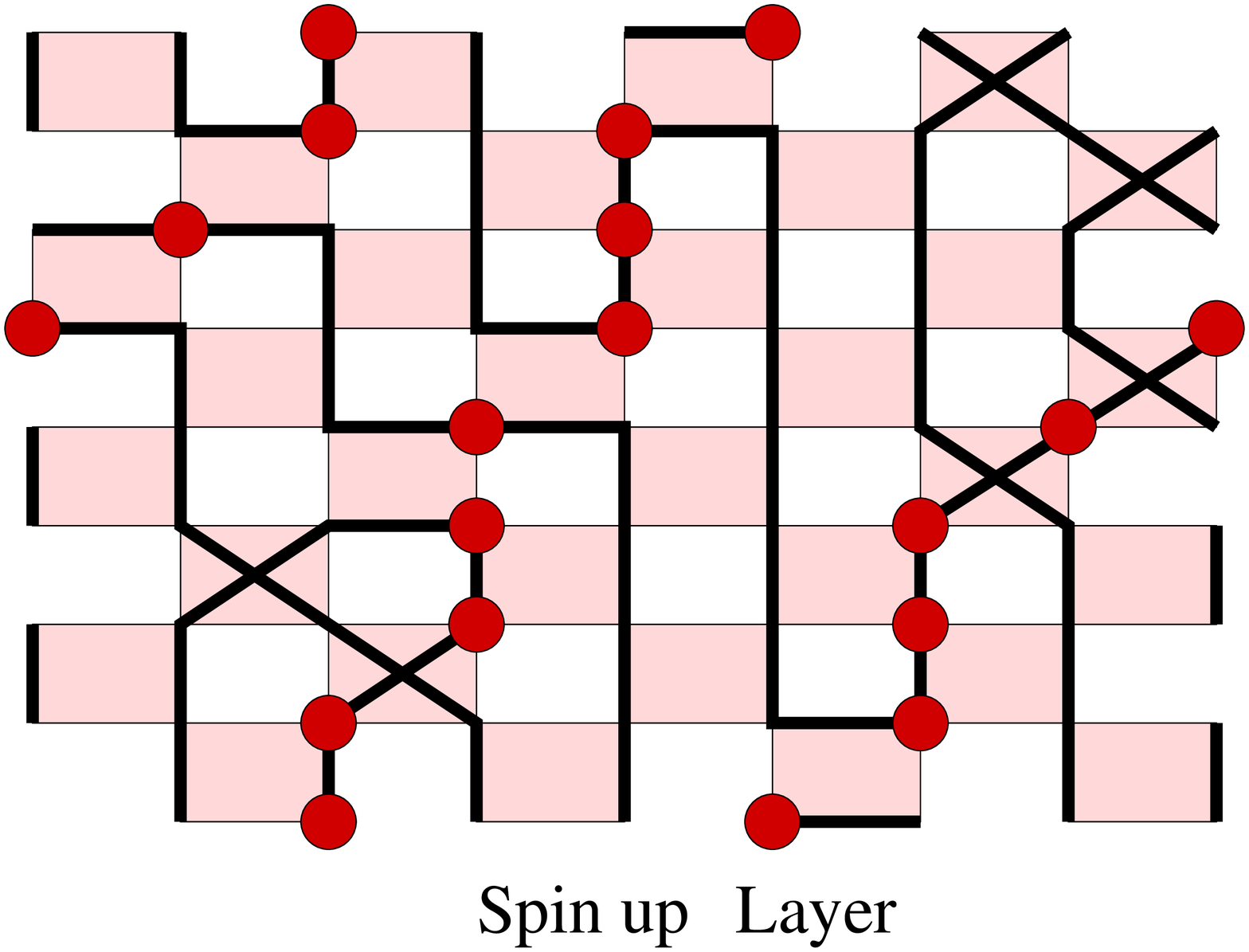}
\hskip0.1in
\includegraphics[width=0.48\textwidth]{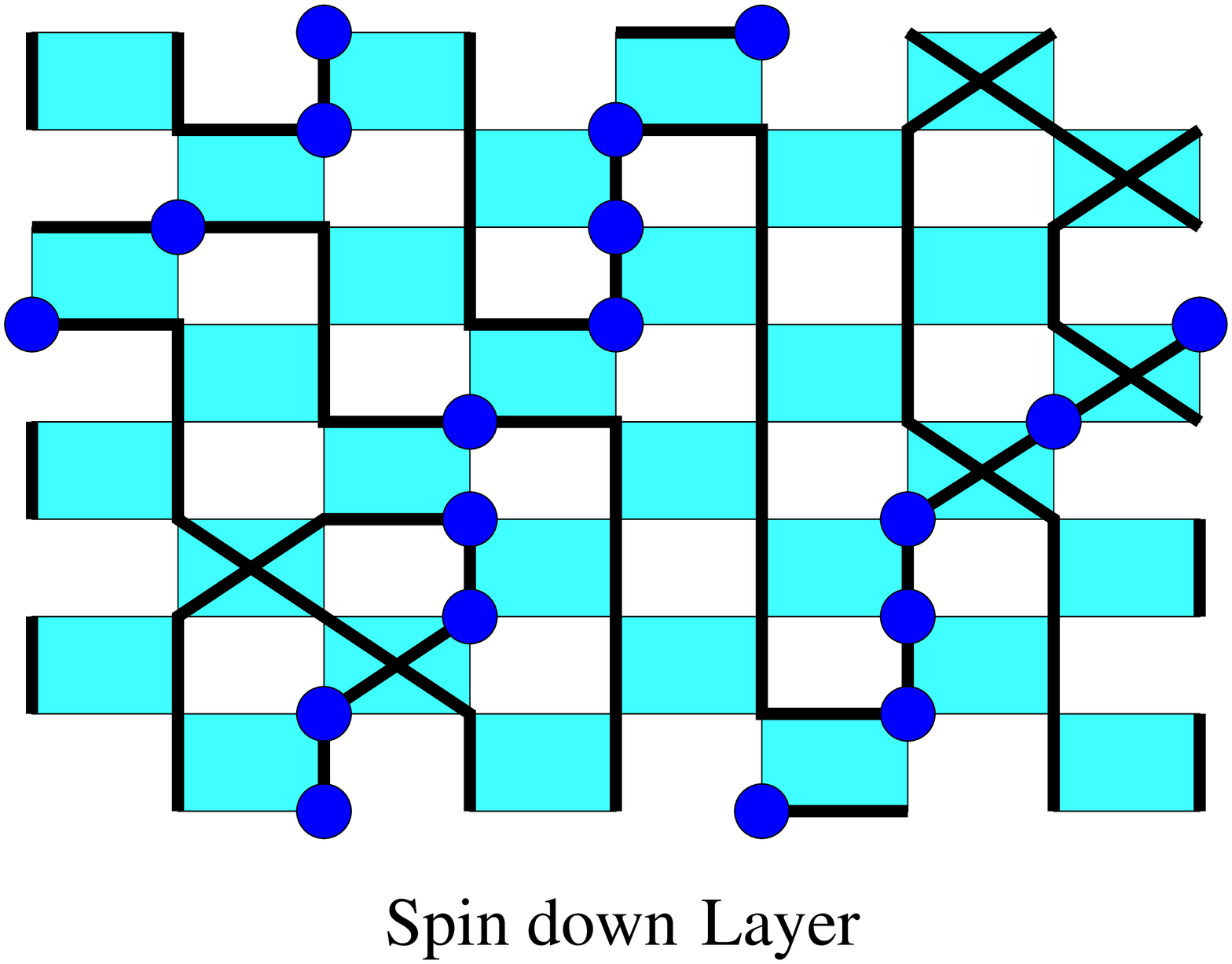}
}
\end{center}
\caption[]{\it A typical reference configuration for the Hubbard-type 
model described in the text.}
\label{spinrefconf}
\end{figure}

Variants of the Hubbard model are equally interesting. For example,
the $t$-$J$ model has also been extensively studied.
Here we refer to such models as Hubbard-type models. Let us now construct
one such model which is solvable using meron-cluster techniques. The model 
is based on the four-site transfer matrix elements $E$ and $H$ shown in
figure \ref{4sbldblks}. In this model the clusters are identical in the 
spin-up and the spin-down layers and any configuration of clusters can be 
flipped into a configuration which is identical in the two layers. Such
a configuration always has a positive sign and can hence play the role of a
reference configuration. A typical reference configuration is shown 
in figure \ref{spinrefconf}. Interestingly, unlike in the spinless fermion 
case, the reference configuration is non-static and not unique. This leads 
to interesting physical consequences. Another feature of the model is that 
there are always an even number of meron-clusters: if the cluster in 
the spin-up layer is a meron then so is the cluster in the spin-down layer.

Using the operators of table \ref{tab4sbldblks} the nearest-neighbor Hamilton 
operator for the model turns out to be
\begin{eqnarray}
h_{xy}&=&- [{\cxd}^\dagger \cyd + {\cyd}^\dagger \cxd - 
2(\nxd - 1/2)(\nyd-1/2) + 1/2] \nonumber \\
&\times&[{\cxu}^\dagger \cyu + {\cyu}^\dagger \cxu - 
2(\nxu - 1/2)(\nyu-1/2) + 1/2] \nonumber \\
&-&\Delta [{\cxd}^\dagger \cyd + {\cyd}^\dagger \cxd - \nxd - \nyd + 1]
\nonumber \\
&\times&[{\cxu}^\dagger \cyu + {\cyu}^\dagger \cxu - \nxu - \nyu + 1].
\label{rawh}
\end{eqnarray}
which leads to the Hamilton operator $H = \sum_{<xy>} h_{xy}$. The
operator given in eq.(\ref{rawh}) can be simplified to \cite{Osb00}
\begin{eqnarray}
h_{xy}&=&({c_{x s}}^\dagger c_{y s} + {c_{y s}}^\dagger c_{x s})
(n_{xy} - 1)(n_{xy} - 3) + 
2 (\vec{S}_x\cdot\vec{S}_y +\vec{J}_x\cdot\vec{J}_y) \nonumber \\
&-&4 \left(\nxu -\frac{1}{2}\right)\left(\nxd -\frac{1}{2}\right)
\left(\nyu -\frac{1}{2}\right)\left(\nyd -\frac{1}{2}\right)
\nonumber \\
&+&\Delta \left[
({c_{x s}}^\dagger c_{y s} + {c_{y s}}^\dagger c_{x s})(n_{xy} - 2) +
2 (\vec{S}_x\cdot\vec{S}_y +\vec{J}_x\cdot\vec{J}_y) - 4 J^3_x J^3_y\right.
\nonumber \\
&-&\left.\left(\nxu -\frac{1}{2}\right)\left(\nxd -\frac{1}{2}\right) -
\left(\nyu -\frac{1}{2}\right)\left(\nyd -\frac{1}{2}\right)\right],
\label{hubbh}
\end{eqnarray}
where we have used the definition $n_{xy} = \nxu + \nxd + \nyu + \nyd$.
The operators $\vec{S}_x$ and $\vec{J}_x$ are given by the expressions
that depend on the site $x$ inside the summation signs in eq.(\ref{sgen})
and (\ref{jgen}).

The symmetries of the Hamilton operator given in eq.(\ref{hubbh}) can 
easily be found. It is interesting to note that the model is invariant 
under the $SU(2)$ spin symmetry and the $U(1)$ fermion number symmetry. 
Further for $\Delta=0$ the $U(1)$ symmetry enhances to the full $SU(2)$ 
pseudo-spin symmetry. Although unconventional, the physics of this model
is closely related to that of the usual attractive Hubbard model.
In particular, it was found recently that this model shows
s-wave superconductivity even at half-filling \cite{Cha01}.
Later we will see that it is possible to add a chemical potential 
and an on-site attractive interaction to the above Hamiltonian without
introducing sign problems. Further, when $\Delta=0$, it is 
possible to include an on-site repulsion between the fermions of opposite 
spin even in the presence of a chemical potential as long as a certain 
inequality
is satisfied. It is interesting to investigate if this ``repulsive'' model 
can be a useful candidate for high-$T_c$ superconductivity.

\subsection{Improved Estimators}
\label{is2}

Since the model satisfies all the properties of the meron-cluster approach,
its partition function can be expressed purely in terms of clusters,
\begin{equation}
Z \;=\; \sum_{[b],{\mathrm{zero-meron}}}\; 
2^{2N_{\cal C}}\; W[b],
\label{clpf2}
\end{equation}
where $N_C$ is the number of clusters on each spin layer. Since the
clusters are identical we get an extra factor of two as compared to
eq.(\ref{clpf1}). This novel way of re-writing the partition function
involves a partial re-summation over cluster-flips and can be extended
to other observables as well. We discussed how the deviation of
the occupation number from half-filling can be expressed in this new 
language in section \ref{is1}. 

Let us consider observables relevant for s-wave superconductivity.
An important observable is the s-wave pair susceptibility
\begin{equation}
P_L \; = \;
 \frac1{\beta V Z} \; \int_0^\beta \;dt\;
 {\mbox Tr} \left[ \; \exp[-(\beta-t) H]\; (p^+ + p^-) \;
 \exp[-t H]\; (p^+ + p^-) \;\right]
\label{pairsus}
\end{equation}
with
$p^+ = \sum_x c_{x \uparrow}^\dagger c_{x \downarrow}^\dagger$
the pair creation and $p^- = (p^+)^\dagger$ the pair annihilation 
operators. An improved estimator for this susceptibility is easy to 
construct using the results of \cite{Bro98}. In particular, one can 
show that non-zero 
contributions to the two-point correlations arise only if the creation 
and annihilation operators occur on the same cluster. Further, in
the case of the pair susceptibility it is also important that the clusters 
these operators reside on have been flipped to be identical. The contribution 
is then proportional to the sum of all possible creation and 
annihilation points along the cluster. This just gives a factor 
of the size of the cluster squared. Averaging over the different possible 
cluster-flips gives for the zero-meron sector
\begin{equation}
\left< P_L \right>_{\rm{cluster-flips, zero-meron}} =
  \frac1{N_t^2 V} \sum_{\cal C} \frac12 |{\cal C}|^2,
\end{equation}
where $|{\cal C}|$ is the size of cluster ${\cal C}$. In the two-meron sector 
we get
\begin{equation}
\left< P_L \right>_{\rm{cluster-flips, two-meron}} =
  \frac1{N_t^2 V} \frac12 |{\cal C}_{\mathrm{meron}}|^2.
\end{equation}
Note that in the two-meron sector there are two identical
meron-clusters: one in the spin-up and one in the spin-down layer.

Another observable relevant to superconductivity is the helicity modulus
\cite{helmod,Har98} which is defined in terms of the winding number as
\begin{equation}
\Upsilon = \frac{T}{8} \langle W_x^2 + W_y^2 \rangle.
\label{fnwind}
\end{equation}
Here $W_x$ ($W_y$) is the total number of fermions winding around the spatial
boundary in the $x$- ($y$-) direction. To measure $W_x$ we need to examine 
each interaction plaquette crossing the boundary in the $x$-direction.
We count +1 for each fermion (up or down spin) that hops across the 
plaquette in the forward direction, $-1$ for each that hops in reverse and 
0 otherwise. This winding for a plaquette can then be easily expressed in 
terms of the occupation numbers of the sites touching the plaquette.
Consider first just the up spin layer.  The winding number on a plaquette that
originates from the point ${x,t}$ and goes in the forward $i$-direction for one
spin layer is
\begin{equation}
W_{x,t,i} = \frac12 \left( n_{x,t} + n_{x+\hat{i},t+1} - n_{x+\hat{i},t} -
        n_{x,t+1} \right) ~.
\end{equation}
We can also replace each $n$ above with $n-1/2$ without changing the results
to make the values more symmetric. Now every site counts as $\pm1/4$ 
(on a single spin layer) depending on its occupation number and where
it is located on the boundary plaquette.
Each cluster can then be assigned a spatial winding number which is
just the sum of
the weights associated with each point on the cluster.
Flipping a cluster inverts all the occupation numbers and thus changes the
sign of the spatial winding number for that cluster. The total winding number
in the $x$-direction is then a sum over all clusters and both spin layers
\begin{equation}
W_x = \sum_{\cal C} W_x^{{\cal C} \uparrow} + W_x^{{\cal C} \downarrow} ~.
\end{equation}
After squaring this we average over all cluster-flips.
In the zero-meron sector one obtains
\begin{equation}
\langle W_x^2 \rangle_{\rm{cluster-flips, zero-meron}} =
\sum_{\cal C} (W_x^{{\cal C} \uparrow})^2 + (W_x^{{\cal C} \downarrow})^2 = 
2 \sum_{\cal C} (W_x^{{\cal C} \uparrow})^2.
\end{equation}
In the two-meron sector we get
\begin{equation}
\langle W_x^2 \rangle_{\rm{cluster-flips, two-meron}} =
2 |W_x^{{\cal C}_{\mathrm{meron}} \uparrow}|
|W_x^{{\cal C}_{\mathrm{meron}} \downarrow}|
= 2 (W_x^{{\cal_C}_{\mathrm{meron}} \uparrow})^2.
\end{equation}
Again, the two-meron sector consists of one meron-cluster in the 
spin-up layer and one meron-cluster in the spin-down layer both of which 
are identical.

Using similar procedures it is also possible to calculate other observables
like the spin susceptibility which is important to detect magnetic 
transitions, the d-wave pair susceptibility which is important for high-$T_c$ 
superconductivity, and various other two-point correlation functions
which are of interest. In principle, it is also possible to calculate
higher-point correlation functions. However, these will become increasingly
noisy and would require more statistics.

\subsection{Kosterlitz-Thouless Transitions}

As discussed above, the attractive Hubbard model has been studied extensively
as a toy model for s-wave superconductivity \cite{Mor91,Lac98}. 
It is expected that below a critical temperature transportation of fermion 
number through the bulk becomes easy, leading to superconductivity (or more 
appropriately superfluidity since the symmetry is not gauged in the model). 
In higher dimensions this is related to the spontaneous breaking of the 
$U(1)$ fermion number symmetry. In two dimensions, since this is forbidden 
due to the Mermin-Wagner theorem, superconductivity occurs due to the 
Kosterlitz-Thouless (K-T) phenomena \cite{KT72}.
\begin{figure}[ht]
\vskip0.3in
\begin{center}
\includegraphics[width=0.9\textwidth]{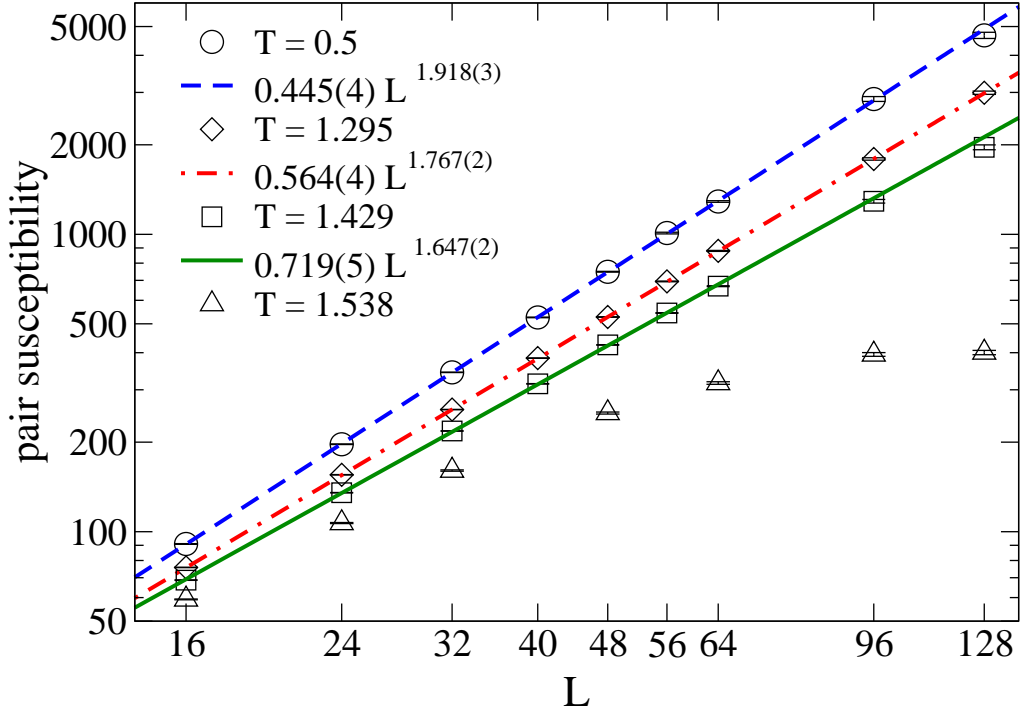}
\end{center}
\caption{\label{psus}
Pair susceptibility as a function of $L$.}
\end{figure}

A variety of striking predictions exist for a K-T transition. For example, 
at temperatures above the superconducting transition temperature $T_c$, the 
pairing susceptibility defined in eq.(\ref{pairsus}) should reach a 
constant for large enough volumes.  As $T_c$ is approached from above this 
constant should diverge as $\mathrm{exp}(a/\sqrt{T-T_c})$. Below $T_c$ one 
should be in a phase with long-range correlations such that the 
finite-volume pair susceptibility diverges as
\begin{equation}
P_L \;\propto \; L^{2-\eta(T)}.
\label{fsspl}
\end{equation}
The critical exponent $\eta$ equals $1/4$ at $T_c$ and continuously changes
to zero as the temperature approaches zero. Unfortunately, none of these 
have ever been confirmed using conventional algorithms even in the 
attractive Hubbard model which does not suffer from a sign problem
in the conventional formulation. 

We propose that the model given in eq.(\ref{hubbh}) is useful for
studying the qualitative physics of the attractive Hubbard model since
it provides a computational advantage. To illustrate this,
recently the finite temperature phase transition in the model with 
$\Delta = 1$ was studied \cite{Cha01}. In this case, the Hamilton operator 
can be re-written as
\begin{eqnarray}
H &=& \sum_{<xy>}\;\Bigl[
\sum_{s={\uparrow,\downarrow}}
({c_{x,s}}^\dagger c_{y,s} + {c_{y,s}}^\dagger c_{x,s})
(1 - 3 n_{xy} + n_{xy}^2)
\nonumber \\
-&4&\left(\nxu -\frac{1}{2}\right)\left(\nxd -\frac{1}{2}\right)
\left(\nyu -\frac{1}{2}\right)\left(\nyd -\frac{1}{2}\right)
\nonumber \\
+&4& \left.
[\vec{S}_x\cdot\vec{S}_y +\vec{J}_x\cdot\vec{J}_y - J^3_x J^3_y]
\right]
\nonumber \\
-&4&\sum_x\;
\;\left(\nxu -\frac{1}{2}\right)\left(\nxd -\frac{1}{2}\right).
\label{hubbkt}
\end{eqnarray}
The spin symmetry and the conservation of fermion number is obvious. 
Further, there is an on-site attraction between electrons of opposite 
spins like in the attractive Hubbard model. In figure 
\ref{psus} we show results for the pair susceptibility as a function of 
the spatial size $L$ which demonstrate that the transition
in this model indeed belongs to the K-T universality class.

The helicity modulus is also a useful observable to establish the K-T
universality class. It can be shown that it satisfies the finite size 
scaling form
\begin{equation}
\frac{\pi}{T}\Upsilon = 2 + \sqrt{A(T)} \coth
\left(\sqrt{A(T)} \log(L/L_0(T)) \right),
\label{fsshm}
\end{equation}
with $A(T_c)=0$ \cite{Har98}. The helicity modulus has also been measured 
using the improved estimator discussed earlier and the results further
confirm the predictions of Kosterlitz and Thouless. More details can be 
found in Ref. \cite{Cha01}.

\subsection{Inclusion of a Chemical Potential}

Until now we have only considered models which can be written in terms
of clusters that satisfy the three requirements necessary for the
meron-cluster
techniques to be applicable. However, in certain cases it is 
possible to relax these stringent requirements and construct models 
which still have no sign problem. Let us now demonstrate this by extending
the Hubbard-type model discussed in section \ref{hubbmodel}. We add a 
single-site term of the form
\begin{equation}
H' = \sum_x \left\{U \left(\nxu - \frac{1}{2}\right)
\left(\nxd - \frac{1}{2}\right) + \mu \left(\nxu + \nxd - 1\right)\right\}.
\end{equation}
It is easy to see that $U$ induces an on-site repulsion or attraction like 
in the Hubbard model and $\mu$ is just a chemical potential. We have already
discussed that in the absence of this single-site term, the partition 
function of the model is given by
\begin{equation}
Z = \sum_{[b],{\mathrm{zero-meron}}} 2^{2N_{\cal C}} W[b],
\end{equation}
where the origin of $2^{2N_{\cal C}}$ can be traced to the four flips of
the two identical clusters in the spin-up and the spin-down layer. Further,
the reference configuration, obtained when both clusters are flipped 
identical to each other is guaranteed to be positive. It is straightforward
to argue that the above single-site term modifies the
factor $2^{2N_{\cal C}}$ to
\begin{equation}
\prod_{{\cal C}}\;
\left\{2\;
\exp\left[-\frac{\epsilon}{2d}\;\frac{U}{4}\;S_{{\cal C}}\right]
\cosh\left(\frac{\epsilon}{2d}\;\mu \Omega_{\cal C}\right) \pm 
2\;\exp\left[\frac{\epsilon}{2d}\;\frac{U}{4}\;S_{{\cal C}}\right] \right\}
\label{sfactor}
\end{equation}
where $S_{{\cal C}}$ is the size of the cluster and $\Omega_{\cal C}$ 
is the number of time-slices times the temporal winding number of the 
cluster. The negative sign should be taken for a meron-cluster.

For the sign problem to remain solved it is important that the factor 
given in eq.(\ref{sfactor}) remains positive. Clearly, this is the
case when $U \leq 0$ for any $\mu$. Interestingly, since the size of
the cluster $S_{\cal C}$ is always greater than or equal to $\Omega_c$,
it is possible to show that even when $U > 0$ and $\mu < U/2$ the factor
remains positive. This means that we have found a repulsive Hubbard-type
model in which it is possible to add a limited chemical potential 
before the sign problem kicks in. Although it is likely that the interesting 
physics of high-$T_c$ superconductivity is beyond this region, it appears 
worthwhile to study this novel model in this region of parameter space
where the sign problem remains solved. It is likely that something 
interesting can be learned there. Finally, in the $U\rightarrow\infty$ 
limit one recovers the Heisenberg anti-ferromagnet. Anti-ferromagnetism
is known to be a part of the high-$T_c$ phase diagram.

\section{Efficiency of the Meron-Cluster Algorithm}

Representing the partition function in terms of the statistical mechanics of
clusters interacting locally has the obvious advantage that numerical
simulations can become efficient. In this section we compare the
efficiency of the meron-cluster algorithm with that of conventional
fermion algorithms.

The efficiency of a numerical simulation method can be characterized by the
computational cost (the computer time $\tau$) that one must invest in order to 
reach a given numerical accuracy. Obviously, the cost depends on the 
$d$-dimensional spatial volume $V = L^d$, where $L$ is the number of lattice 
points per direction. In addition, one usually discretizes the Euclidean 
time-direction of extent $\beta = M \epsilon$ into $M$ time-steps of size 
$\epsilon$, such that the computational cost increases as one approaches 
the time continuum limit $\epsilon \rightarrow 0$. Besides these two
obvious dependences, $\tau$ typically also depends on the spatial correlation 
length $\xi$ via a dynamical exponent $z$ that characterizes the severity 
of critical slowing down. Further, even if $\xi$ is small, the correlation 
length $\xi_t/\epsilon$ in the time-direction becomes large in units of 
the temporal lattice spacing $\epsilon$ when one approaches the time 
continuum limit. Hence, there is another dynamical critical exponent $z_t$ 
that characterizes the corresponding slowing down. Altogether, the total 
computational cost is typically given by
\begin{equation}
\tau \propto L^x (\beta/\epsilon)^y [c \xi^z + c' (\xi_t/\epsilon)^{z_t}].
\end{equation}

Standard local algorithms for bosonic systems --- for example, the Metropolis 
algorithm --- have $z, z_t \approx 2$. This makes it difficult to simulate 
systems close to a critical point, i.e.~for large $\xi$, or close to the time 
continuum limit, i.e.~for small $\epsilon$. Local over-relaxation algorithms 
can be fine-tuned to reach $z, z_t \approx 1$. Efficient non-local cluster 
algorithms, on the other hand, are far superior because they can reach 
$z, z_t \approx 0$. Algorithms for bosonic systems typically require a 
computational effort that increases linearly with the space-time volume, 
i.e.~$x = d$ and $y = 1$. For simulations of fermionic systems this is 
typically not the case, even when there is no sign problem. Usually, one 
integrates out the fermions and simulates a bosonic theory with the fermion 
determinant defining a non-local effective action. 
The standard fermion algorithm 
that is most popular in simulations of strongly correlated electron systems 
\cite{Whi89} computes the determinant of a large matrix of a size proportional 
to the spatial volume. This requires a computational effort proportional to the
spatial volume cubed, i.e.~$x = 3d$, even when there is no sign problem, and 
hence
\begin{equation}
\tau \propto L^{3d} (\beta/\epsilon) [c \xi^2 + c' (\xi_t/\epsilon)^2].
\end{equation}
A modification of this algorithm proposed in \cite{Saw98} has $x = 2d$. It 
should be noted that fermion algorithms that have originally been developed 
for QCD simulations have also been applied to strongly correlated electron 
systems without a sign problem. The Hybrid Monte Carlo algorithm \cite{Dua87}
that was used in \cite{Lac98} has
\begin{equation}
\tau \propto L^{5d/4} (\beta/\epsilon)^{5/4} [c \xi + c' \xi_t/\epsilon],
\end{equation}
and the multi-boson algorithm \cite{Lue94} used in \cite{Saw98} has
\begin{equation}
\tau \propto L^d (\beta/\epsilon) [c \xi + c' \xi_t/\epsilon].
\end{equation}
Here we have taken the optimistic point of view that one can, at least in
principle, reach $z, z_t \approx 1$ by fine-tuning these algorithms. Their 
computational cost has a much better large volume behavior than the standard 
algorithm, but they are not necessarily superior on the presently studied 
moderate size lattices.

When there is a sign problem, the computational cost of the above algorithms
increases even exponentially --- not just as a power --- in the space-time 
volume. As discussed before,
\begin{equation}
\tau \propto \exp(2 \beta V \Delta f),
\end{equation}
such that formally $x = y = \infty$. In practice, this means that
systems with a severe sign problem can simply not be simulated. The
meron-cluster algorithm allows us to deal with the sign problem
extremely efficiently. In particular, the computational cost increases
only linearly with the space-time volume, such that $x = d$ and $y =
1$. Still, this does not necessarily mean that we can reach $z \approx 0$ as in
cluster simulations of bosonic models. This is due to the re-weighting
step in the meron-cluster algorithm. In this step the clusters are
locally reconnected such that multi-meron configurations are never
generated.  For example, a non-meron-cluster may be decomposed into
two meron-clusters only if this does not increase the total meron
number beyond two. Deciding if a newly generated cluster is a meron
naively requires a computational effort proportional to $\xi$, such
that $z \approx 1$ for the meron-cluster algorithm.
However, with recent improvements in the implementation we have been able to
reduce this effort to only grow with $\log(\xi)$.
A nice feature of the meron-cluster algorithm is that --- like other
cluster algorithms --- it can be implemented directly in the Euclidean time 
continuum \cite{Bea96}. This completely eliminates all time discretization 
errors and all $\epsilon$-factors in the computational cost. The total computer
time that is required to reach a given numerical accuracy with the 
meron-cluster algorithm is hence given by
\begin{equation}
\tau \propto L^d \beta [c \xi^z + c' (\xi_t/\epsilon)^{z_t}],
\end{equation}
and it is likely that $z, z_t \approx 0$. This is better than standard methods 
even when there is no sign problem, and exponentially better than any 
alternative method when there is a severe sign problem.

\section{Directions for the Future}

In this article we have outlined the essential steps that lead to a
novel formulation of fermionic lattice theories. The essential
idea behind the new approach is to find a cluster representation of
the partition function. We have shown that in this approach new models
can be formulated for which the sign problem can be completely eliminated.
Further, although the clusters themselves are non-local they interact
locally and carry information about two-point correlations. This makes them
especially attractive for numerical simulations. Given the difficulty
in the numerical treatment of fermionic theories, this alternative
approach turns out to be very useful.

There are many interesting questions that remain unanswered in the
case of strongly correlated electron systems. Among these the
physics of competing ground states that may exist in
high-$T_c$ materials as the doping concentration is increased is
one of the most challenging. The model we have constructed
appears to have many interesting features that are relevant for
this type of questions. This includes anti-ferromagnetism and s-wave
superconductivity at the two extremes of the parameter ranges. It
would be interesting if this model shows d-wave superconductivity
or striped phases in some intermediate range of parameters. Most
importantly, the new Hubbard-type Hamilton operator is well suited 
for numerical work in a large region of parameter space. It would be 
interesting to study this model in this range. As a first step a mean 
field analysis may be useful. 

The superconductor-insulator transition is another interesting field
of research that the present progress may shed some light on \cite{Gol98}.
The attractive Hubbard model with disorder has recently been used to 
study this type of quantum phase transition \cite{Sca99}. We believe 
that the Hamilton operator of eq.(\ref{hubbh}) with disorder is
an alternative model for this purpose. This is possible because the
sign problem remains solved in the presence of disorder. Further, since
the algorithm for the new model is expected to be much more efficient, 
it may be possible to go to much smaller temperatures.

It would be interesting to find extensions of the present 
work to fields like nuclear physics and to nano-science applications
like quantum dots. It is easy to construct models with four layers 
representing spin and isospin which naturally describe the physics 
of nucleons. It would be useful to construct these explicitly so that 
we can be sure that the relevant symmetries can be maintained. 
Extensions of the cluster techniques to models in the continuum is 
another useful direction. Preliminary work shows that the 
non-relativistic free fermion model of eq.(\ref{freef}) can be extended 
to the continuum. It would be interesting if this approach can be extended 
to other interacting non-relativistic free fermions. Such extensions can
be useful in studying physics of many electrons enclosed within boundaries
as in quantum dots.

\section*{Acknowledgments}

This work is supported in part by funds provided by the U.S. Department 
of Energy (D.O.E.) under cooperative research agreements DE-FC02-94ER40818 
and DE-FG-96ER40945, the National Science Foundation under the
agreement DMR-0103003 and the European Community's Human Potential Programme
under HPRN-CT-2000-00145 Hadrons/Lattice QCD, BBW Nr. 99.0143.

\end{document}